\documentclass[onecolumn,showpacs,preprintnumbers,elsart]{revtex4}
\usepackage{amsmath}
\usepackage{mathrsfs}
\usepackage{graphicx}
\usepackage{dcolumn}
\usepackage{bm}
\usepackage[center]{subfigure}
\usepackage{color}

\begin{document}

\title{Nonlinear modes and symmetry breaking in rotating double-well
potentials}
\author{Yongyao Li$^{1,2}$, Wei Pang$^{3}$, and Boris A. Malomed$^{1}$}
\email{malomed@post.tau.ac.il}
\affiliation{$^{1}$Department of Physical Electronics, School of Electrical Engineering,
Faculty of Engineering, Tel Aviv University, Tel Aviv 69978, Israel\\
$^{2}$Department of Applied Physics, South China Agricultural University,
Guangzhou 510642, China \\
$^{3}$ Department of Experiment Teaching, Guangdong University of
Technology, Guangzhou 510006, China.}

\begin{abstract}
We study modes trapped in a rotating ring carrying the self-focusing (SF) or
defocusing (SDF) cubic nonlinearity and double-well potential $\cos
^{2}\theta $, where $\theta $ is the angular coordinate. The model, based on
the nonlinear Schr\"{o}dinger (NLS) equation in the rotating reference
frame, describes the light propagation in a twisted pipe waveguide, as well
as in other optical settings, and also a Bose-Einstein condensate (BEC)
trapped in a torus and dragged by the rotating potential. In the SF and SDF
regimes, five and four trapped modes of different symmetries are found,
respectively. The shapes and stability of the modes, and transitions between
them are studied in the first rotational Brillouin zone. In the SF regime,
two symmetry-breaking transitions are found, of subcritical and
supercritical types. In the SDF regime, an antisymmetry-breaking transition
occurs. Ground-states are identified in both the SF and SDF systems.
\end{abstract}

\pacs{42.65.Tg; 03.75.Lm; 47.20.Ky; 05.45.Yv}
\maketitle




\section{Introduction}

The concept of the spontaneous symmetry breaking (SSB) in nonlinear systems
was introduced in Ref. \cite{Chris}. Its significance has been later
recognized in various physical settings, including numerous ones originating
in nonlinear optics \cite{Snyder}-\cite{photo}, Bose-Einstein condensates
(BECs) \cite{Milburn}-\cite{Arik}, and degenerate fermionic gases \cite%
{Padua}. A general analysis of the SSB phenomenology was developed too \cite%
{misc}, which is closely related to the theory of bifurcations in nonlinear
systems \cite{Bif}.

Fundamental manifestations of the SSB occur in nonlinear systems based on
symmetric double-well potentials (DWPs) or dual-core configurations. A
paradigmatic example of the latter in nonlinear optics is the twin-core
nonlinear fiber, which may serve as a basis for the power-controlled optical
switching \cite{Snyder}. DWP settings in optics were analyzed theoretically
and implemented experimentally in photorefractive crystals \cite{photo}. In
the realm of matter waves, main effects predicted in DWPs are Josephson
oscillations \cite{Zapata}, the asymmetric self-trapping of localized modes
\cite{Warsaw}, and similar effects in binary mixtures \cite{Mazzarella,Ng}.
Both the Josephson and self-trapping regimes were implemented in the atomic
condensate with contact repulsive interactions \cite{Markus}. The SSB was
also analyzed in one- and two-dimensional (1D and 2D) models of BEC trapped
in dual-core configurations \cite{Arik}.

Another dynamical setting which has produced a number of interesting effects
in media with the intrinsic nonlinearity, especially in BEC, is provided by
rotating potentials. It is well known that stirring the self-repulsive
condensate typically leads to the formation of vortex lattices \cite%
{vort-latt}, although it was found experimentally \cite{Cornell} and
demonstrated theoretically \cite{giant} that giant vortices, rather
than lattices, may also be formed under special conditions (when the
centrifugal force nearly compensates the trapping
harmonic-oscillator potential). On the other hand, the rotation of
self-attractive condensates gives rise to several varieties of
stable localized modes, such as vortices, ``crescents"
(mixed-vorticity states), and the so-called center-of-mass modes
(quasi-solitons) \cite{rotating-trap}. Further development of this
topic was achieved by the consideration of rotating lattice
potentials, which can be implemented experimentally as an optical
lattice induced in BEC by a broad laser beam transmitted through a
revolving sieve \cite{sieve}, or using \textit{twisted}
photonic-crystal fibers in optics \cite{twistedPC}. In these
systems, quantum BEC states and vortex lattices have been studied
\cite{in-sieve}, as well as solitons and solitary vortices depinning
from the lattice when its rotation velocity exceeds a critical value
\cite{HS}.

A specific implementation of the latter settings is provided by the
quasi-1D lattice, or a single quasi-1D potential well, revolving
about its center in the 2D geometry \cite{Barcelona}. In particular,
the rotation makes it possible to create fundamental and vortical
soliton in the self-repulsive medium, where, obviously, nonrotating
quasi-1D potentials cannot maintain bright solitons \cite{Kon}.
Furthermore, the rotation of a DWP gives rise to azimuthal Bloch
bands \cite{Ueda,Stringari}.

\begin{figure}[tbp]
\centering{\includegraphics[scale=0.6]{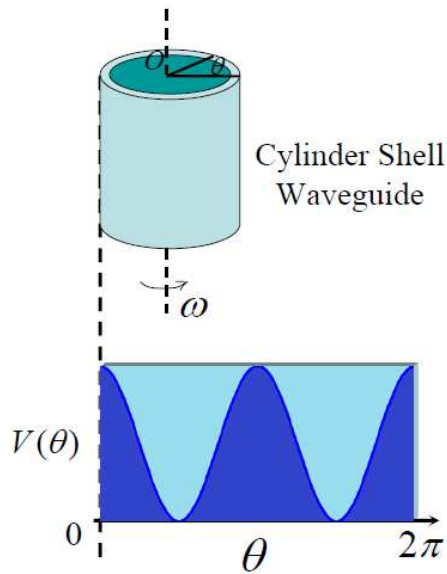}}
\caption{(Color online) The pipe waveguide with the intrinsic potential,
twisted at rate $\protect\omega $.}
\label{fig_1}
\end{figure}

As mentioned above, the static DWP and its limit form reducing to dual-core
systems are fundamental settings for the onset of the SSB \cite{Snyder}-\cite%
{Arik}. A natural problem, which is the subject of the present work, is the
SSB and related phenomenology, i.e., the existence and stability of
symmetric, antisymmetric, and asymmetric modes, in \emph{rotating} DWPs
(recently, a revolving DWP configuration was considered in a different
context in Ref. \cite{WenLuo}, as a stirrer generating vortex lattices). To
analyze basic features of the phenomenology, we here concentrate on the
one-dimensional DWP placed onto a rotating ring. As shown in Fig. \ref{fig_1}%
, in optics this setting may be realized as a hollow pipe waveguide twisted
with pitch $2\pi /\omega $, while the azimuthal modulation of the refractive
index, that gives rise to the effective potential $V(\theta )$, is written
into the material of the pipe. Alternatively, a \textit{helical} potential
structure can be created in a straight sheath waveguide by means of
optical-induction techniques, using pump waves with the ordinary
polarization in a photorefractive material (while the probe wave is to be
launched in the extraordinary polarization \cite{Moti}), or the method of
the electromagnetically-induced transparency (EIT) \cite{Fleischhauer},
including its version recently proposed for supporting spatial solitons \cite%
{Yongyao}. In the latter case, one can make the pipe out of Y$_{2}$SiO$_{5}$
crystal doped by Pr$^{3+}$ (Pr:YSO) ions \cite{Kuznetsova}. In either case
of the use of the photorefractive material or EIT, the helical structure may
be induced by a superposition of a pair of co-propagating \textit{vortical}
pump waves, with equal amplitudes, a small mismatch of the propagation
constants $k_{1,2}$ ($\Delta k\equiv k_{1}-k_{2}\ll k_{1}$), and opposite
vorticities $\left( \pm S\right) $, which will give rise to an effective
potential profile,
\begin{equation}
V(\theta ,z)\sim r^{S}\cos \left( \Delta k\cdot z+2S\theta \right) ,
\label{V}
\end{equation}%
where $z\ $is the propagation distance, while $r$ and $\theta $ are the
polar coordinates in the transverse plane. In terms of the BEC, a similar
setting may be based on ring-shaped (toroidal) traps, which have been
created in experiments \cite{torus} and investigated in various contexts
theoretically \cite{Salasnich}. In that case, the rotating periodic
potential can be added to the toroidal trap \cite{sieve} , which is
equivalent to the consideration of the rotating ring \cite%
{Stringari,rotating-ring}.

In this work, we study basic types of trapped modes and their SSB
phenomenology in the 1D rotating ring, in both cases of the self-focusing
and self-defocusing (SF and SDF) cubic nonlinearities. In Sec. II we
formulate the model and present analytical results, which predict a boundary
between the symmetric and asymmetric modes, the analysis being possible for
the small-amplitude potential and the rotation rate close to $\omega =1/2$.
Numerical results are reported in a systematic form, and are compared to the
analytical predictions, in Secs. III and IV for the SF and SDF
nonlinearities, respectively. The paper is concluded by Sec. IV.

\section{The model and analytical considerations}

As said above, we consider the limit of a thin helical shell, which implies
a fixed value of the radius in Eq. (\ref{V}), $r=r_{0}$, that we normalize
to be $r_{0}=1$. Taking the harmonic periodic potential in the form of Eq. (%
\ref{V}) with $S=1$, $V(\theta ,z)=2A\cos ^{2}(\theta -\omega z)$, the
corresponding scaled nonlinear Schr\"{o}dinger equation is
\begin{equation}
i{\frac{\partial }{\partial z}}\psi =\left[ -{\frac{1}{2}}{\frac{\partial
^{2}}{\partial \theta ^{2}}}+V(\theta ,z)-\sigma |\psi |^{2}\right] \psi ,
\label{Eq5}
\end{equation}%
where $\sigma =+1$ and $-1$ refer to SF and SDF nonlinearities,
respectively. Then, we rewrite Eq. (\ref{Eq5}) in the helical coordinate
system, with $\theta ^{\prime }\equiv \theta -\omega z$:
\begin{equation}
i{\frac{\partial }{\partial z}}\psi =\left[ -{\frac{1}{2}}{\frac{\partial
^{2}}{\partial \theta ^{\prime }{}^{2}}}+i\omega {\frac{\partial }{\partial
\theta ^{\prime }}}+2A\cos ^{2}(\theta ^{\prime })-\sigma |\psi |^{2}\right]
\psi ,  \label{Eq5p}
\end{equation}%
where the solution domain is defined at $-\pi \leq \theta ^{\prime }\leq
+\pi $. For the narrow toroidal BEC trap with the rotating potential, the
respective Gross-Pitaevskii equation, written in the co-rotating reference
frame (cf. Refs. \cite{Ueda,Stringari}), differs by replacing the
propagation distance, $z$, with time $t$ \cite{review}.

Stationary modes with real propagation constant $-\mu $ (in terms of the
BEC, $\mu $ is the chemical potential) are sought for as $\psi \left( \theta
^{\prime },z\right) =\exp \left( -i\mu z\right) \phi (\theta ^{\prime })$,
with complex function $\phi \left( \theta ^{\prime }\right) $ obeying
equation
\begin{equation}
\mu \phi =\left[ -{\frac{1}{2}}{\frac{d^{2}}{d\theta ^{\prime }{}^{2}}}%
+i\omega {\frac{d}{d\theta ^{\prime }}}+2A\cos ^{2}(\theta ^{\prime
})-\sigma |\phi |^{2}\right] \phi .  \label{phi}
\end{equation}%
Equation (\ref{Eq5p}) conserves the total power (norm) of the field and its
Hamiltonian (energy),
\begin{equation}
P=\int_{-\pi }^{+\pi }\left\vert \psi (\theta ^{\prime })\right\vert
^{2}d\theta ^{\prime },  \label{Power}
\end{equation}%
\begin{equation}
H=\int_{-\pi }^{+\pi }\left[ {\frac{1}{2}}\left\vert {\frac{\partial \psi }{%
\partial \theta ^{\prime }}}\right\vert ^{2}+{\frac{i}{2}}\omega \left( \psi
^{\ast }{\frac{\partial \psi }{\partial \theta ^{\prime }}}-\psi {\frac{%
\partial \psi ^{\ast }}{\partial \theta ^{\prime }}}\right) +V(\theta
^{\prime })|\psi |^{2}-{\frac{\sigma }{2}}|\psi |^{4}\right] d\theta
^{\prime },  \label{Ham}
\end{equation}%
with the asterisk stands for the complex conjugate.

The periodic boundary conditions, $V(\theta ^{\prime }+2\pi )=V(\theta
^{\prime })$ and $\psi (\theta ^{\prime }+2\pi )=\psi (\theta ^{\prime })$,
make Eq. (\ref{Eq5p}) invariant with respect to the \textit{boost
transformation}, which allows one to change the rotation speed from $\omega $
to $\omega -N$ with arbitrary integer $N$:%
\begin{equation}
\psi \left( \theta ^{\prime },z;\omega -N\right) =\psi \left( \theta
^{\prime },z;\omega \right) \exp \left[ -iN\theta ^{\prime }+i\left( \frac{1%
}{2}{N}^{2}-N\omega \right) z\right] ,  \label{boost}
\end{equation}%
hence the speed may be restricted to interval $0\leq \omega <1$.
Furthermore, Eq. (\ref{Eq5p}) admits an additional invariance, relating
solutions for opposite signs of the rotation speed: $\psi (\theta ^{\prime
},z;\omega )=\psi ^{\ast }(\theta ^{\prime },-z;-\omega )$. If combined with
shift $\omega \rightarrow \omega +1$, the latter transformation implies that
the solutions for the rotation speeds $\omega $ and $1-\omega $ (with $%
0<\omega <1$) are mutually tantamount, therefore the rotation speed may be
eventually restricted to interval
\begin{equation}
0\leq \omega \leq 1/2,  \label{zone}
\end{equation}%
which plays the role similar to that of the fist Brillouin zone in
solid-state media \cite{Kittel}.

An analytical approach for small-amplitude modes can be developed if the
amplitude of the potential in Eq. (\ref{Eq5p}) is small too, $|A|~\ll 1$,
and $\omega $ is close to the right edge of zone (\ref{zone}), $\delta
\equiv 1/2-\omega ~\ll 1/2$. In the simplest approximation, the
corresponding stationary solutions are looked for as a combination of the
zeroth and first angular harmonics,
\begin{equation}
\phi \left( \theta ^{\prime }\right) =a_{0}+a_{1}\exp \left( i\theta
^{\prime }\right) ,  \label{ansatz}
\end{equation}%
where $a_{0}$ is fixed to be real, while amplitude $a_{1}$ may be complex.
Indeed, setting $\omega =1/2$, $A=0$, and dropping the cubic term, it is
obvious that Eq. (\ref{ansatz}) is an exact solution to Eq. (\ref{phi}).
Then, under the aforementioned conditions, the substitution of ansatz (\ref%
{ansatz}) into Eq. (\ref{phi}) yields, in the first approximation (taking
care of the balance of the zeroth and first harmonics in the equation), the
following algebraic equations:
\begin{eqnarray}
&&\mu a_{0}=Aa_{0}-\sigma a_{0}^{3}-2\sigma |a_{1}|^{2}a_{0},  \notag \\
&&\mu a_{1}=\delta a_{1}-\sigma |a_{1}|^{2}a_{1}-2\sigma
a_{0}^{2}a_{1}+Aa_{1}.  \label{Ea0a1}
\end{eqnarray}

First, setting $a_{1}=0$ or $a_{0}=0$ in ansatz (\ref{ansatz}) corresponds,
respectively, to the CW (continuous-wave) mode, and the uniform vortical
one, with $\left\vert \psi \left( \theta ^{\prime }\right) \right\vert =%
\mathrm{const}$:%
\begin{eqnarray}
a_{1} &=&0,~a_{0}^{2}=\sigma \left( A-\mu \right) ,  \label{0} \\
a_{0} &=&0,~\left\vert a_{1}\right\vert ^{2}=\sigma \left( A-\mu +\delta
\right)   \label{1}
\end{eqnarray}%
(recall that $\sigma ^{2}=1$). In terms of full equation (\ref{Eq5p}) with $%
A\neq 0$, these two solutions extend into non-uniform ones [with $\left\vert
\psi (\theta ^{\prime })\right\vert \neq \mathrm{const}$], which remain (and
are categorized as) \textit{symmetric} with respect to point $\theta
^{\prime }=0$. The respective vortical mode, obtained as the extension of
solution (\ref{1}), is investigated in a numerical form in the next section,
under the name of FHS (fundamental-harmonic symmetric) mode.

Solutions to Eqs. (\ref{Ea0a1}) with $a_{0}a_{1}\neq 0$ give rise to the
species of \textit{asymmetric} modes (categorized as FHA, i.e.,
fundamental-harmonic asymmetric mode, in the next section), with
\begin{equation}
a_{0}^{2}={\frac{\sigma }{3}}(A-\mu +2\delta ),~|a_{1}|^{2}={\frac{\sigma }{3%
}}(A-\mu -\delta ).  \label{a0a1}
\end{equation}%
According to Eqs. (\ref{Power}) and (\ref{ansatz}), the propagation constant
is related to total power (\ref{Power}) of soliton (\ref{a0a1}), $P=2\pi
(a_{0}^{2}+|a_{1}|^{2})$, as follows:
\begin{equation}
\mu =A+{\frac{\delta }{2}}-{\frac{3\sigma }{4\pi }}P.  \label{E}
\end{equation}%
In particular, in the case of the SF nonlinearity, with $\sigma =+1$, Eq. (%
\ref{E}) demonstrates that the asymmetric mode meets the
Vakhitov-Kolokolov (VK) criterion, $d\mu /dP<0$, which is a
necessary stability condition for modes supported by the SF terms
\cite{VK}. On the other hand, in the case of the SDF sign of the
nonlinearity, $\sigma =-1$, the asymmetric mode satisfies the
``anti-VK" criterion, $d\mu /dP>0$, which, as argued in Ref.
\cite{anti}, may also play the role of a necessary stability
condition, in the respective setting.

Further, Eq. (\ref{a0a1}) predicts a transition between the symmetric (FHS)
and asymmetric (FHA) vortical modes at $a_{0}^{2}=0$, i.e., at $\mu
=A+2\delta $. Then, Eq. (\ref{E}) yields the location of this boundary in
terms of the total power, $P_{\mathrm{\min }}=2\pi \sigma \delta $. In view
of the above-mentioned equivalence between the modes pertaining to rotation
speeds $\omega =1/2\pm \delta $, the latter result predicts the coexistence
of the FHS and FHA modes at
\begin{equation}
P\geq P_{\mathrm{\min }}=2\pi |\delta |.  \label{P}
\end{equation}%
Comparison of this prediction with numerical findings is presented below.

\section{Numerical results for the \textbf{self-focusing nonlinearity (}$%
\protect\sigma =+1$\textbf{)}}

\subsection{Symmetric, asymmetric, and antisymmetric modes}

Solutions to stationary equation (\ref{phi}) were constructed by
means of numerical code ``PCSOM" elaborated in Ref. \cite{YJK}. In
the SF case, the use of different input waveforms makes it possible
to
identify five distinct species of stationary modes, as listed in Table \ref%
{tab:table1}.

\begin{table}[tbp]
\caption{Different species of stationary modes in the case of the SF
nonlinearity ($\protect\sigma =+1$), labeled by input waveforms which
generate them.}%
\begin{ruledtabular}
\begin{tabular}{lccr}
Inputs  & Types of modes   \\
\hline
 $\cos\theta'$ & Fundamental-harmonic symmetric (FHS)\\
$1+\sin\theta'$ & Fundamental-harmonic asymmetric (FHA)\\
 $\sin^{2}\theta'$ & Second-harmonic symmetric (2HS) \\
$1+\sin\theta'$ & Second-harmonic asymmetric (2HA)\\
$\sin\theta'$ & Anti-symmetric (AnS)\\
$b + \sin\theta',\,0<b\leq 1$ & Broken-antisymmetry (BAnS)
\end{tabular}
\end{ruledtabular}
\label{tab:table1}
\end{table}
The symmetry, asymmetry and antisymmetry of the modes designated in
Table 1 is realized with respect to point $\theta ^{\prime }=0$,
while the solution is considered, as defined above, in the region of
$-\pi \leq \theta ^{\prime }\leq +\pi $. Further, the fundamental or
second\ harmonic (``FH" or ``2H", respectively) in the nomenclature
adopted in the table refers to a dominant term in the Fourier
decomposition of the stationary solution (which is made obvious by
their shapes, see below). In particular, the stationary patterns of
the FHA and 2HA types are generated by the same input in Table 1,
$1+\sin \theta ^{\prime }$, but in non-overlapping regions of the
parameter space, $\left( \omega ,A,P\right) $, as shown below. Note
also that all the inputs displayed in the table are real functions,
but the numerical modes found for $\omega \neq 0$ have a complex
structure.

The stability of the stationary solutions has been identified through the
numerical computation of eigenvalues for infinitesimal perturbation modes.
To this end, the perturbed solution was taken in the usual form,
\begin{equation}
\psi =e^{-i\mu z}[\phi (\theta ^{\prime })+u(\theta ^{\prime })e^{i\lambda
z}+v^{\ast }(\theta ^{\prime })e^{-i\lambda ^{\ast }z}],  \label{pert}
\end{equation}%
where $u(\theta ^{\prime })$ and $v(\theta ^{\prime })$ are perturbation
eigenmodes, and $\lambda $ is the corresponding eigenfrequency. The
substitution of expression (\ref{pert}) into Eq. (\ref{Eq5p}) and
linearization leads to the eigenvalue problem in the following form:
\begin{equation}
\left(
\begin{array}{cc}
\mu {-}\mathrm{\hat{h}}-i\omega \frac{\partial }{\partial \theta ^{\prime }}
& \sigma \phi ^{2} \\
-\sigma \left( \phi ^{\ast }\right) ^{2} & -\mu +\mathrm{\hat{h}}-i\omega
\frac{\partial }{\partial \theta ^{\prime }}%
\end{array}%
\right) \left(
\begin{array}{c}
u \\
v%
\end{array}%
\right) =\lambda \left(
\begin{array}{c}
u \\
v%
\end{array}%
\right) ,  \label{lambda}
\end{equation}%
where $\mathrm{\hat{h}}=-(1/2)\partial ^{2}/\partial \theta ^{\prime
2}+2A\cos ^{2}(\theta ^{\prime })-2\sigma |\phi |^{2}$ is the respective
single-particle Hamiltonian. The underlying solution $\phi $ is stable if
all the eigenvalues are real.

\begin{figure}[tbp]
\centering%
\subfigure[] {\label{fig_2_a}
\includegraphics[scale=0.21]{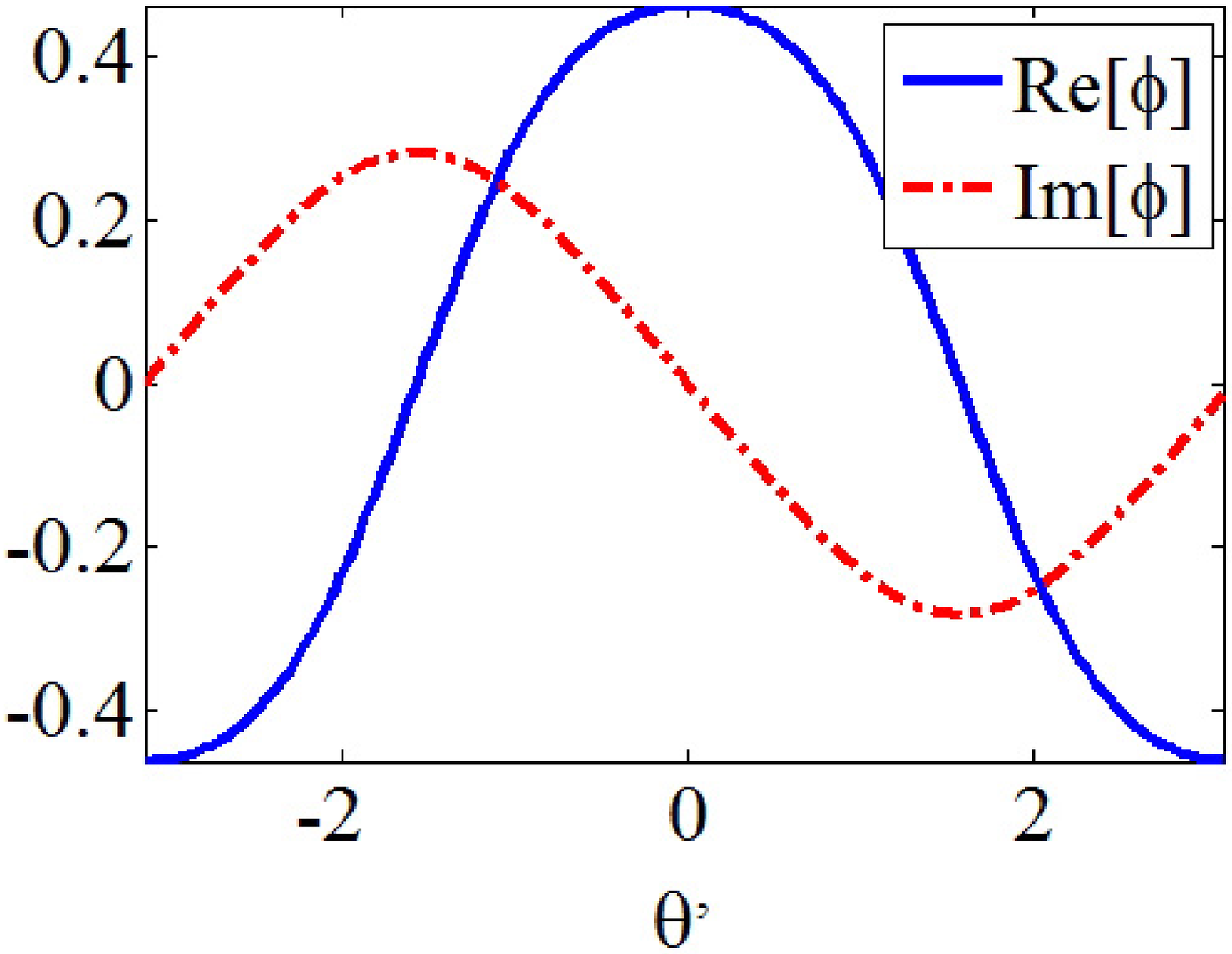}}%
\subfigure[] {\label{fig_2_b}
\includegraphics[scale=0.21]{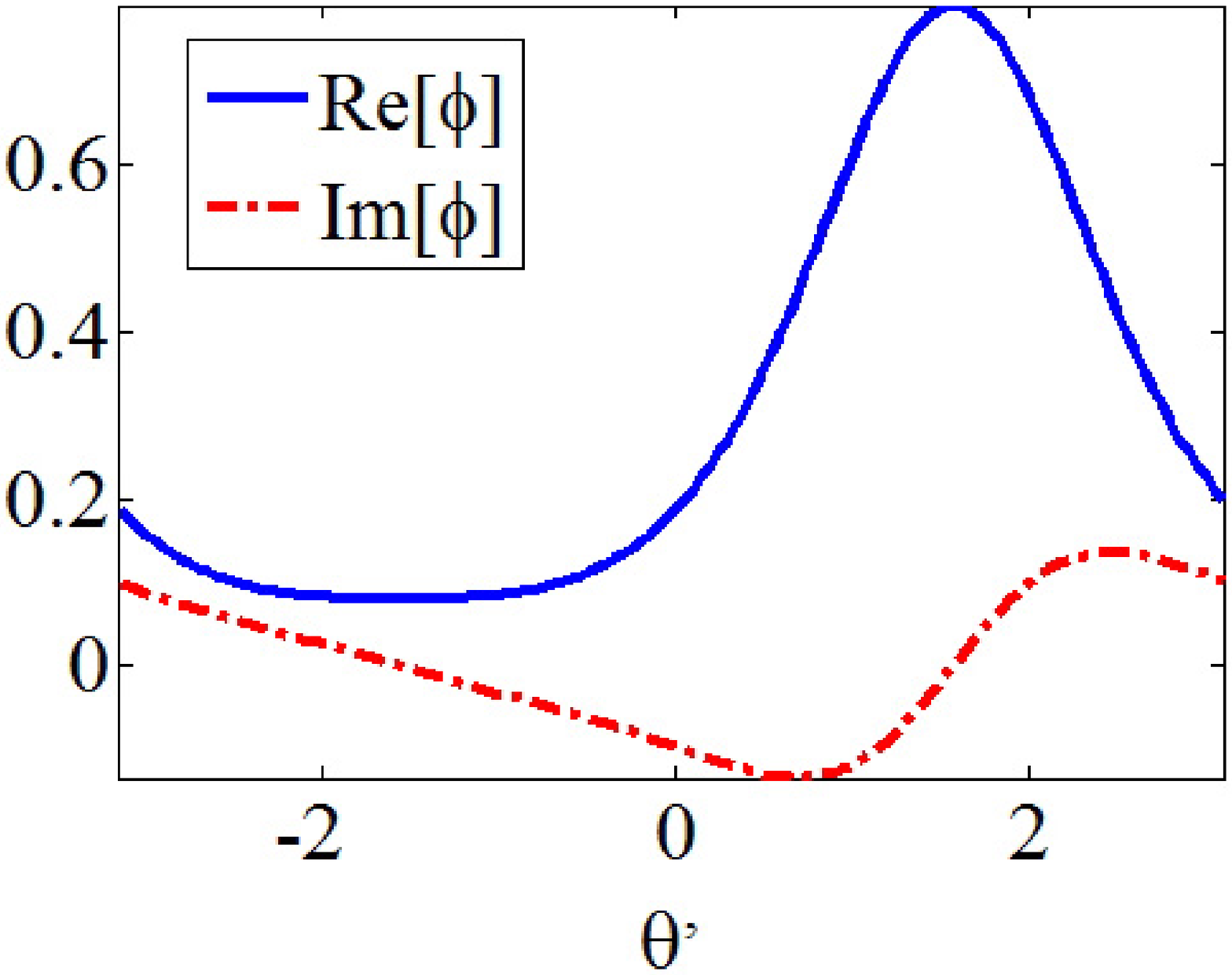}}
\subfigure[]{ \label{fig_2_c}
\includegraphics[scale=0.21]{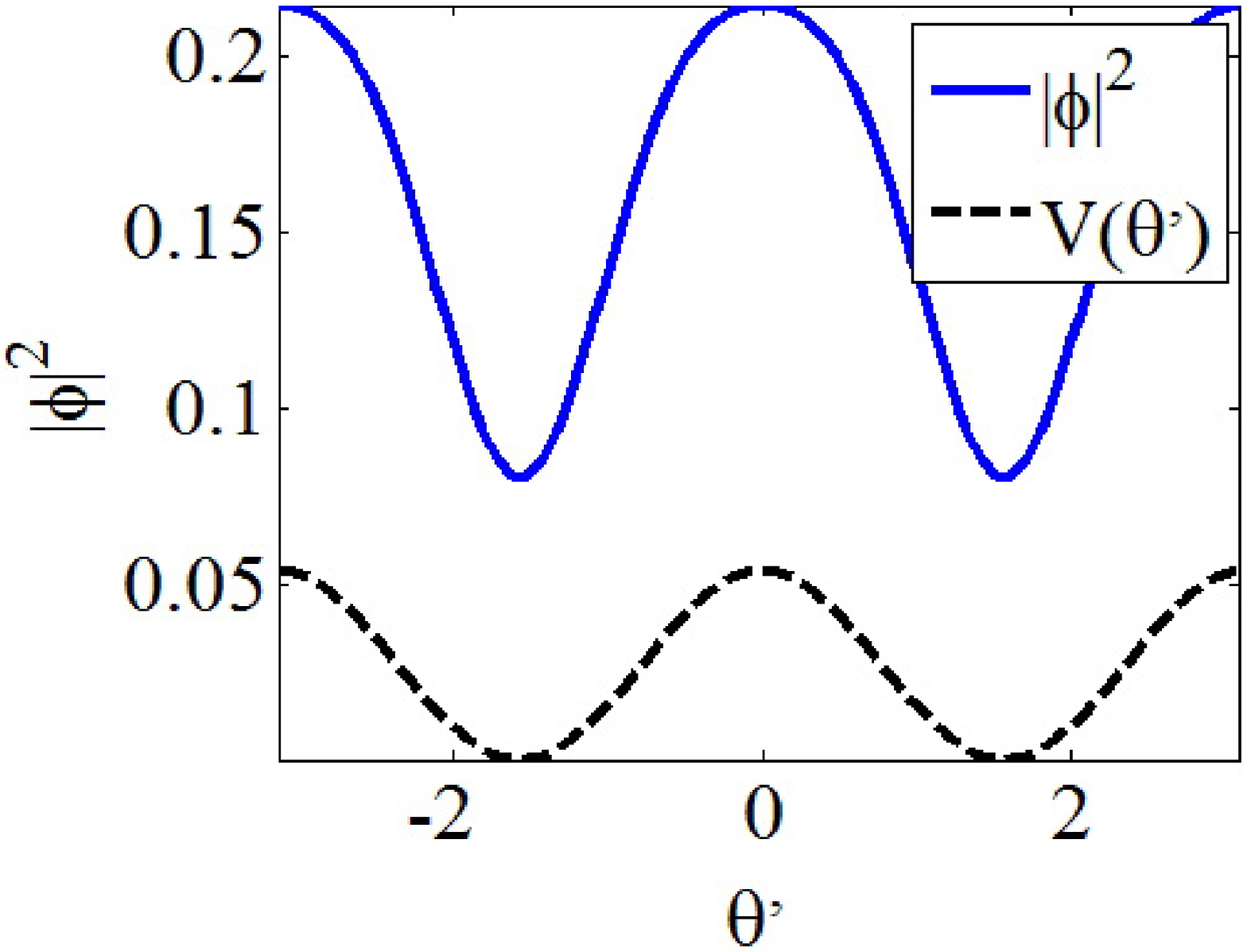}}
\subfigure[]{ \label{fig_2_d}
\includegraphics[scale=0.21]{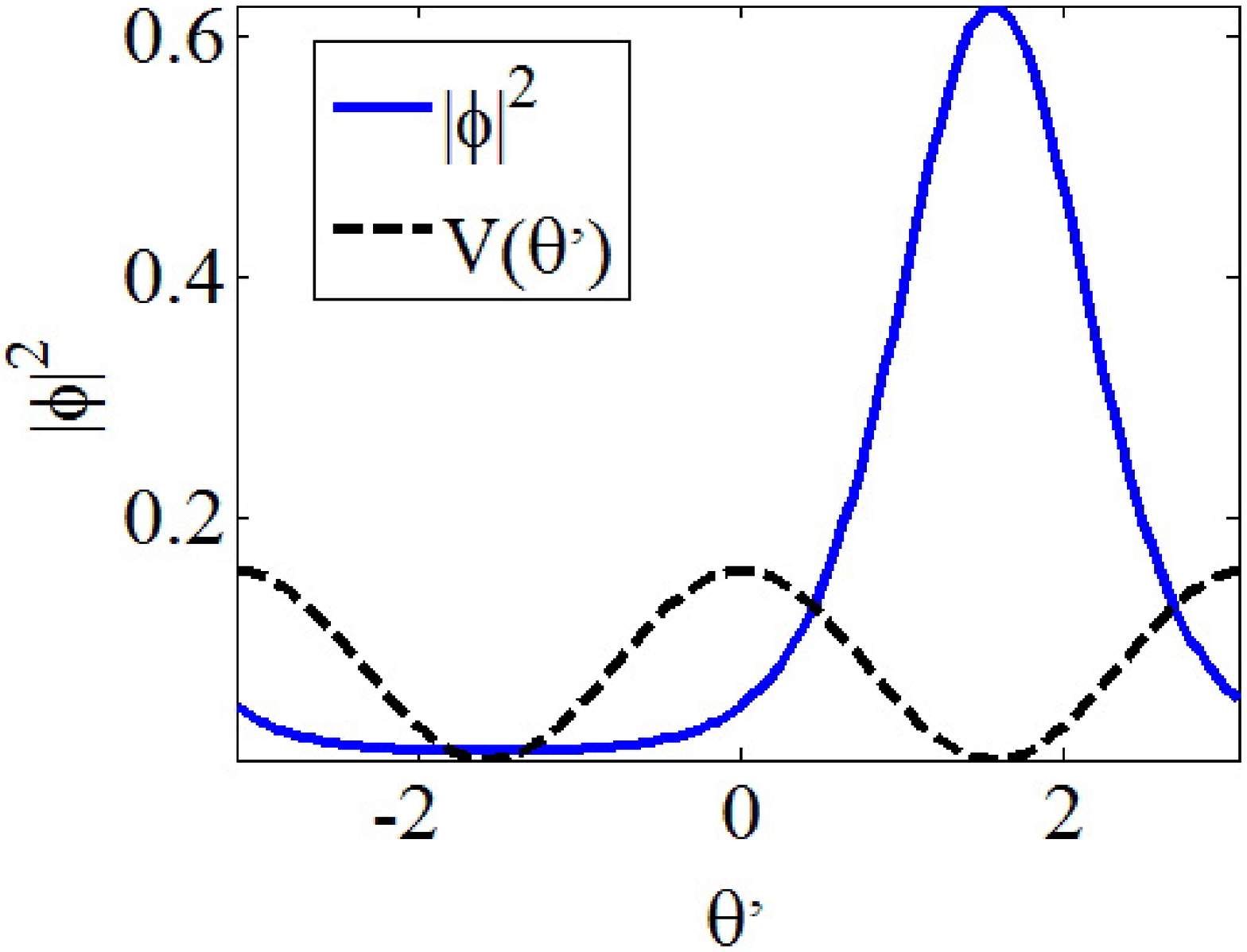}}
\caption{(Color online) Examples of stable FHS and FHA (fundamental-harmonic
symmetric and asymmetric) modes, in the case of the SF nonlinearity,
corresponding to the same parameter set, $(\protect\omega ,A,P)=(0.35,0.5,1)$%
. (a,b) Real and imaginary parts of the FHS and FHA stationary solutions,
respectively. (c,d) Intensity profiles of the same solutions. The dashed
curves in (c) and (d), and in similar panels displayed below, depict the
potential, $V\left( \protect\theta ^{\prime }\right) =2A\cos ^{2}\left(
\protect\theta ^{\prime }\right) $. }
\label{1stSF}
\end{figure}

\begin{figure}[tbp]
\centering%
\subfigure[] {\label{fig_3_a}
\includegraphics[scale=0.21]{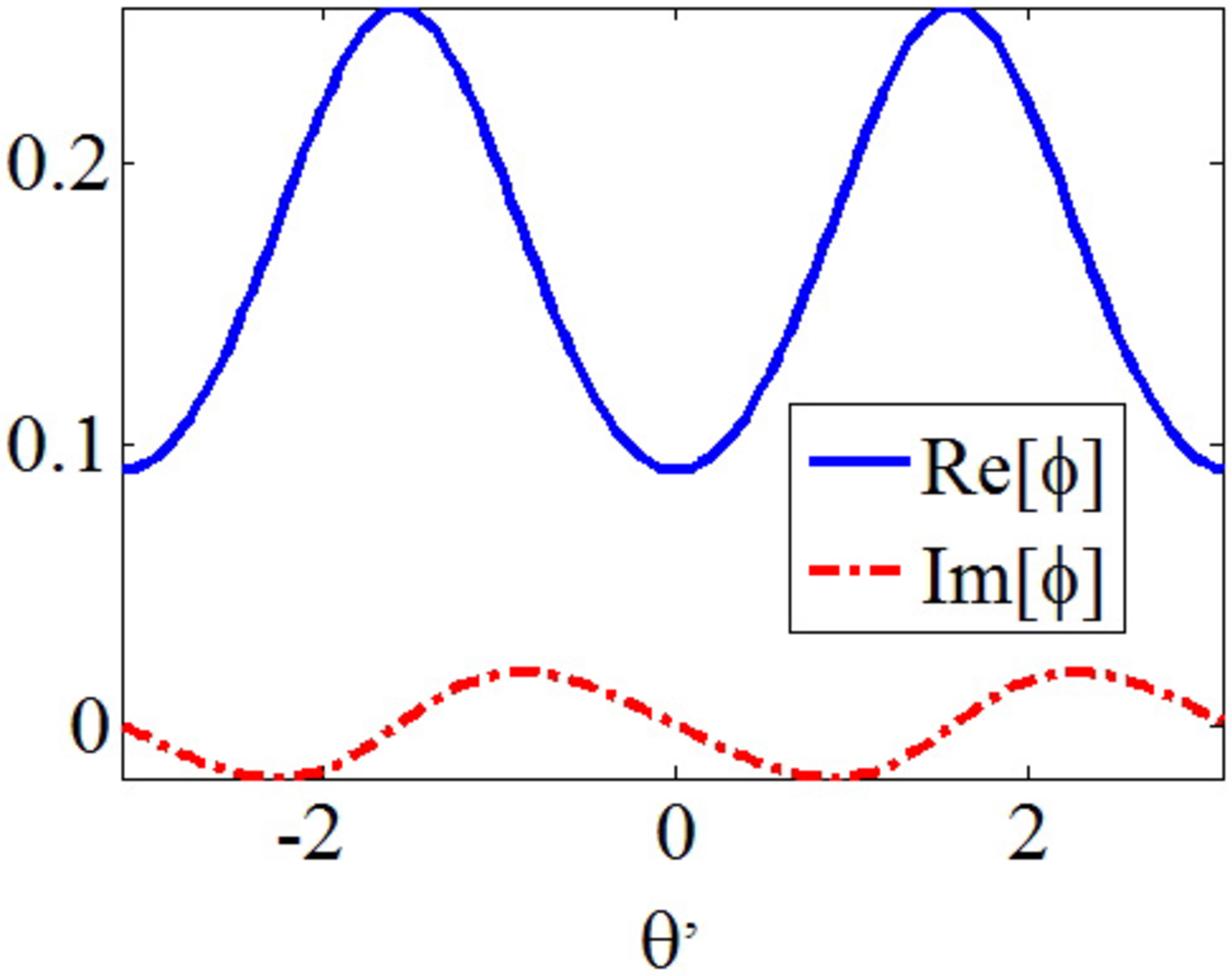}}%
\subfigure[] {\label{fig_3_b}
\includegraphics[scale=0.21]{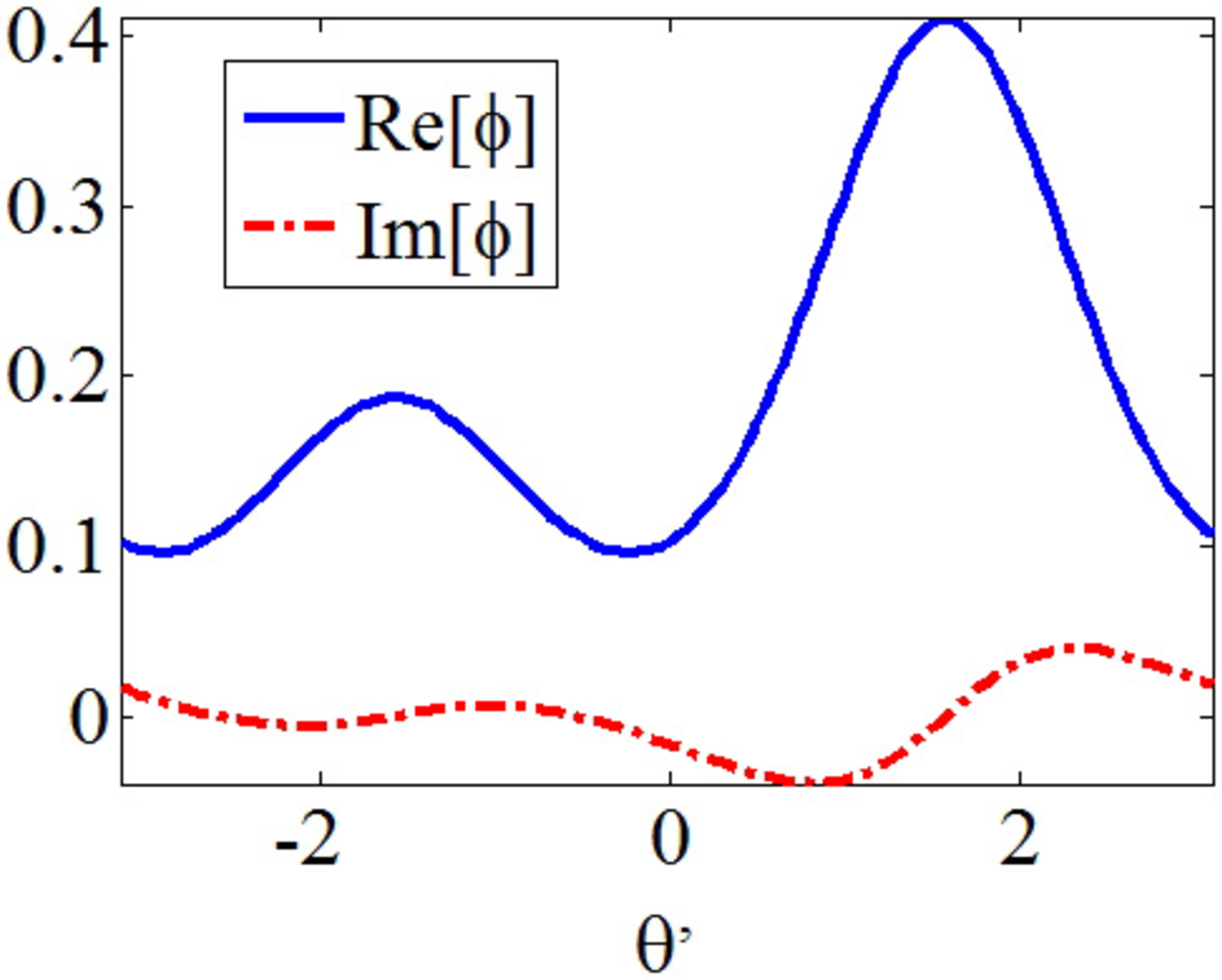}}
\subfigure[]{ \label{fig_3_c}
\includegraphics[scale=0.21]{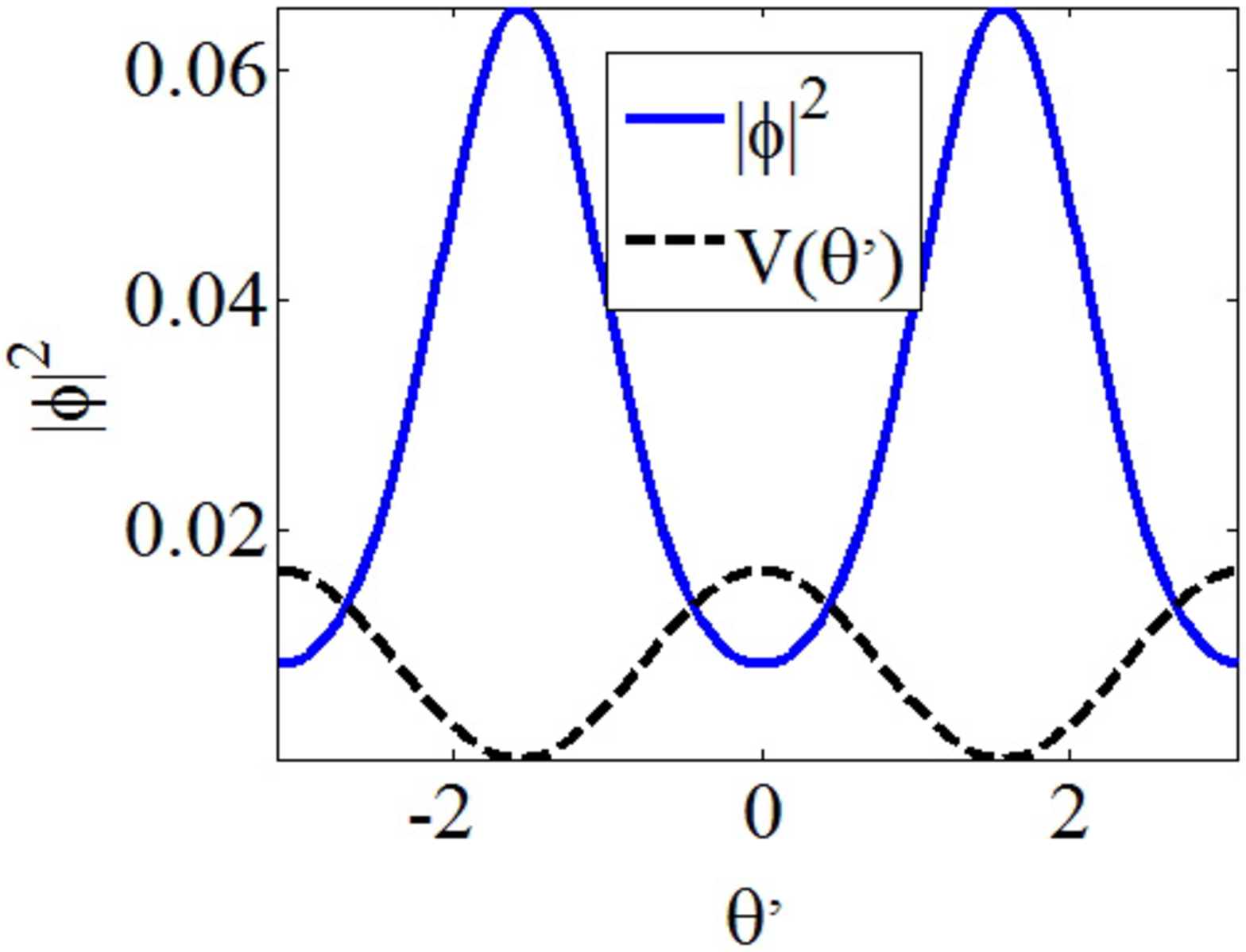}}
\subfigure[]{ \label{fig_3_d}
\includegraphics[scale=0.21]{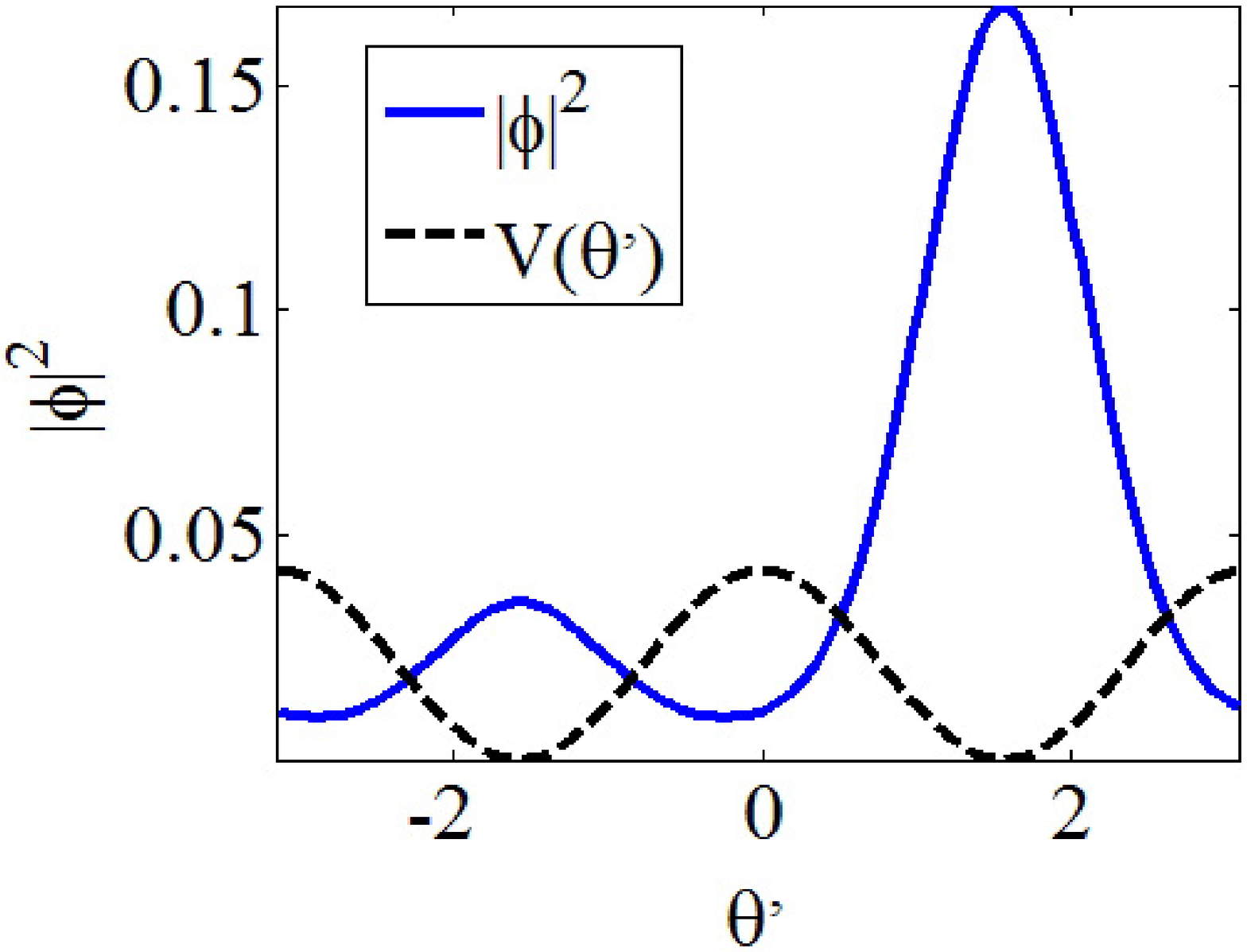}}
\caption{(Color online) Examples of stable modes of the 2HS and 2HA
(second-harmonic symmetric and asymmetric) types, in the case of the SF
nonlinearity, found at sets of the parameter values $(\protect\omega %
,A,P)=(0.25,1,0.2)$ and $(0.25,1,0.3)$, respectively. (a,b) Real and
imaginary parts of the 2HS and 2HA stationary solutions, respectively. (c,d)
Intensity profiles of the same solutions. Unlike the fundamental-harmonic
modes shown in Fig. \protect\ref{1stSF}, these ones, based on the second
harmonic, feature no bistability (coexistence of the stable modes).}
\label{2ndSF}
\end{figure}

\begin{figure}[tbp]
\centering%
\subfigure[] {\label{fig_4_a}
\includegraphics[scale=0.21]{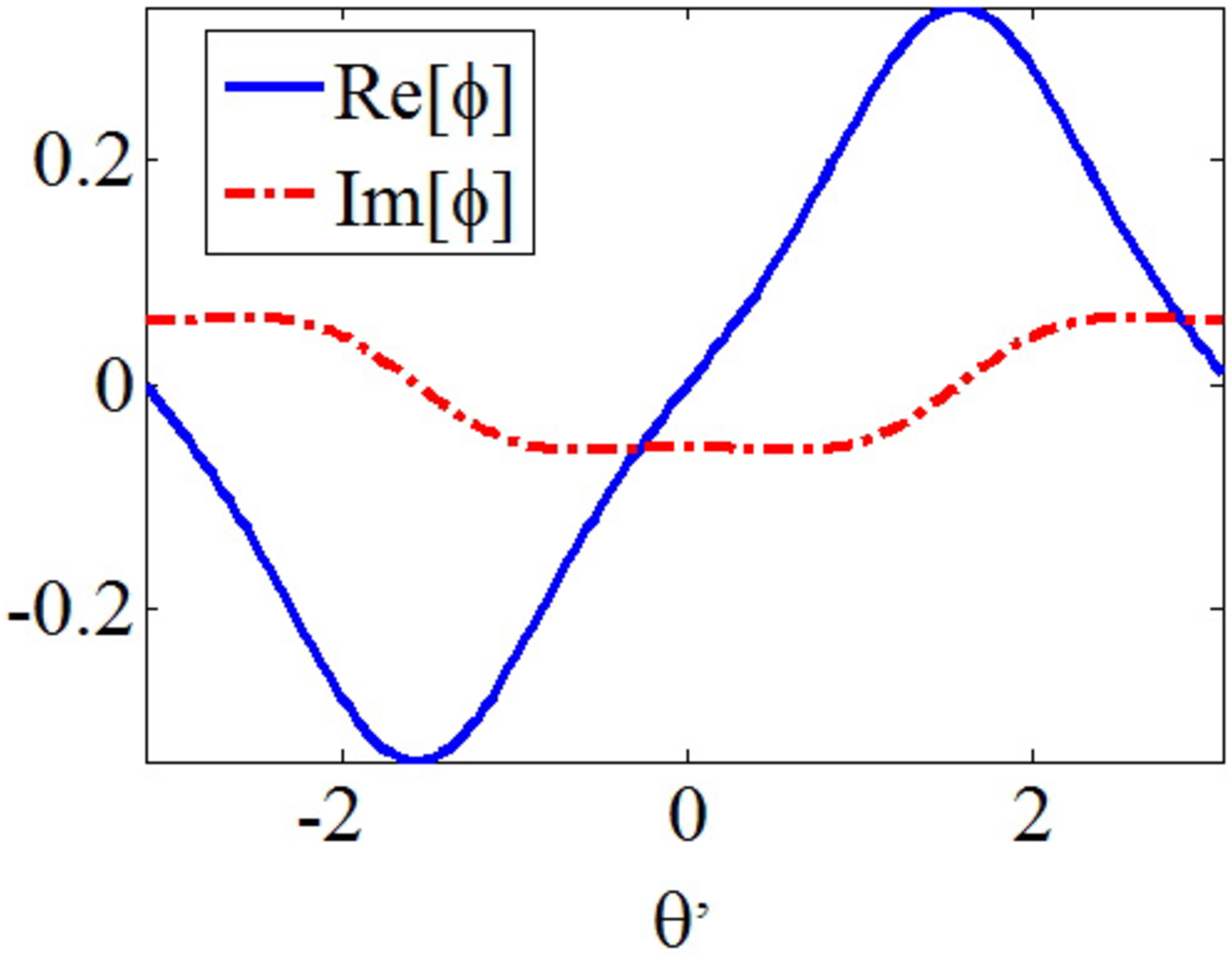}}%
\subfigure[] {\label{fig_4_b}
\includegraphics[scale=0.21]{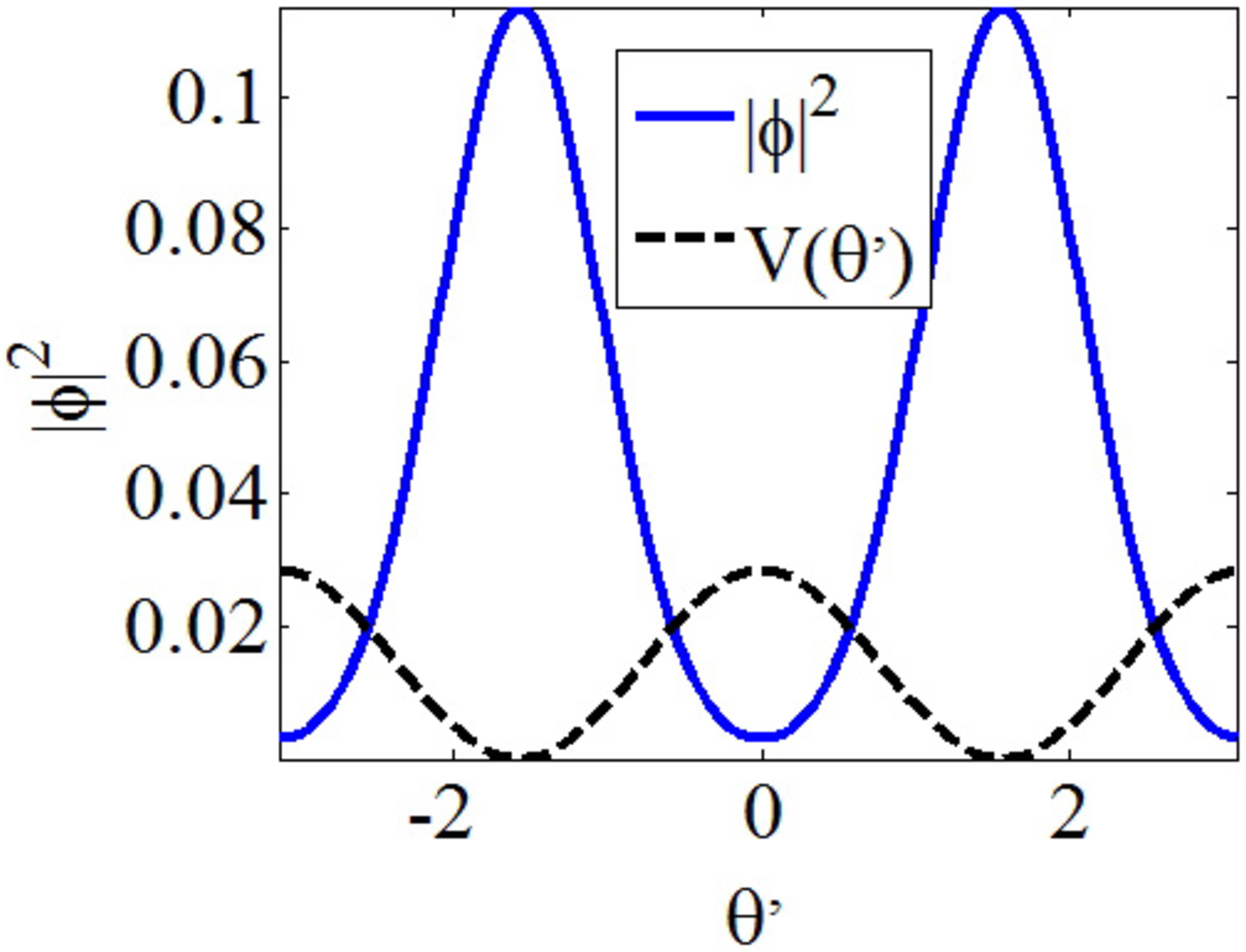}}
\caption{(Color online) An example of a stable mode of the AnS
(antisymmetric) type, at $(\protect\omega ,A,P)=(0.25,1,0.3)$: (a) real and
imaginary parts of the stationary solution; (b) its intensity profile.}
\label{AntiSF}
\end{figure}

In the case of the SF nonlinearity, the numerical solutions reveal the
existence of all the species of the modes indicated in Table 1, except for
BAnS, which exist under the SDF nonlinearity, see below. Typical examples of the five species
of the stationary modes which are supported by the SDF cubic term are
displayed in Fig. \ref{1stSF}-\ref{AntiSF}. The analysis demonstrates that
all these examples are stable. Moreover, the symmetric and asymmetric modes
dominated by the fundamental harmonic, FHS and FHA, demonstrate mutual
\textit{bistability }in Fig. \ref{1stSF}, as these stable modes coexist at
common values of the parameters.

A distinctive feature of the FHS mode, which is evident in Fig. \ref{1stSF},
is that maxima and minima of its intensity coincide with local maxima and
the minima of the harmonic potential, while its FHA counterpart features an
intensity maximum in one potential well, and a minimum in the other. This
structure of the FHS suggests that it may correspond to a maximum, rather
than minimum, of the energy (see below), but, nevertheless, this mode has a
region of stability against small perturbations.

On the other hand, Figs. \ref{2ndSF} and \ref{AntiSF} demonstrates that both
the 2HS and AnS modes have two symmetric intensity peaks trapped in the two
potential wells (while local minima of the intensity coincide, quite
naturally, with potential humps), hence these mode have a chance to realize
a minimum of the energy (in the best case, it may be the system's ground
state). The 2HA mode features a similar property in Fig. \ref{2ndSF}, but
with unequal density peaks trapped in the two potential minima.

\begin{figure}[tbp]
\centering%
\subfigure[] {\label{fig_5_a}
\includegraphics[scale=0.21]{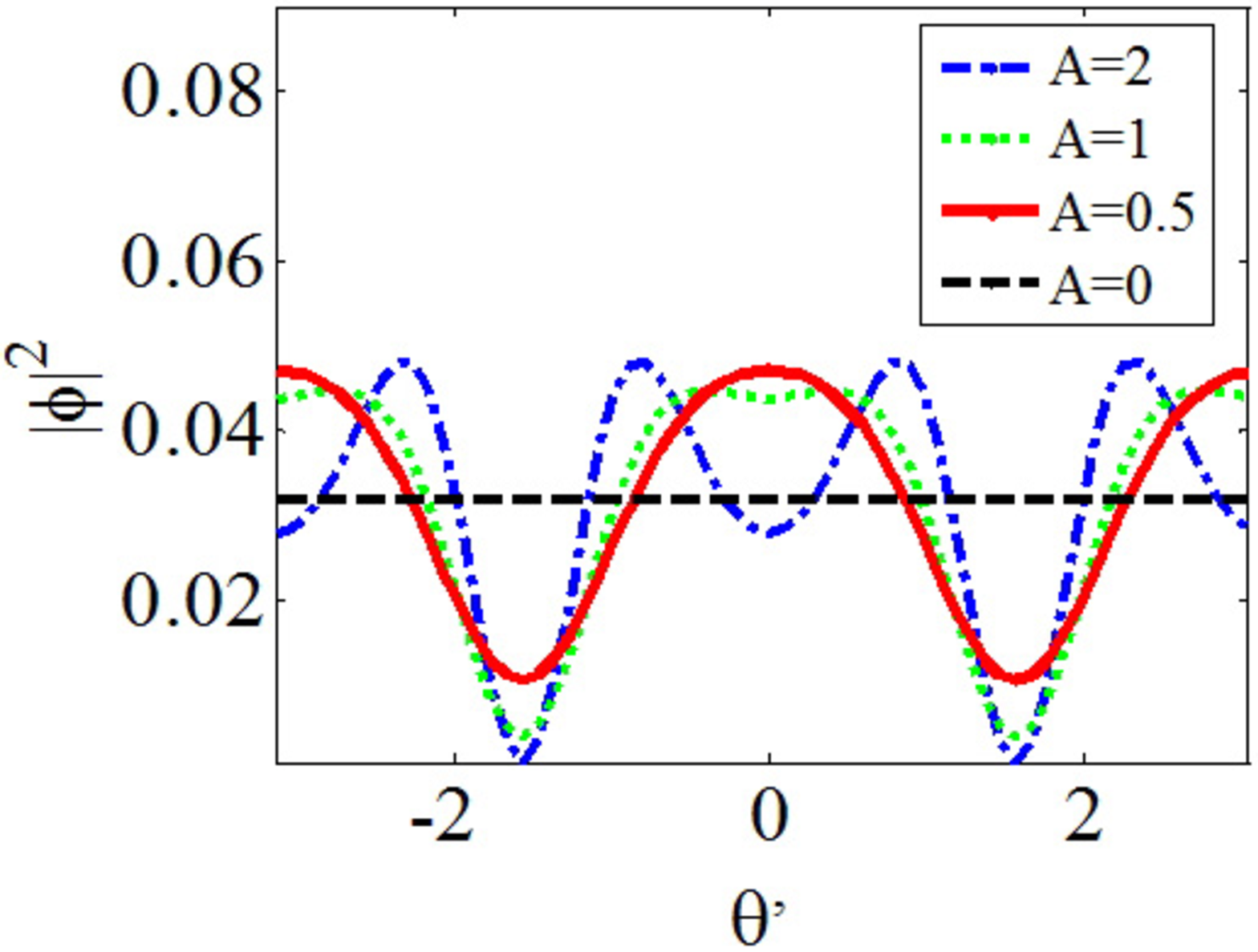}}%
\subfigure[] {\label{fig_5_b}
\includegraphics[scale=0.21]{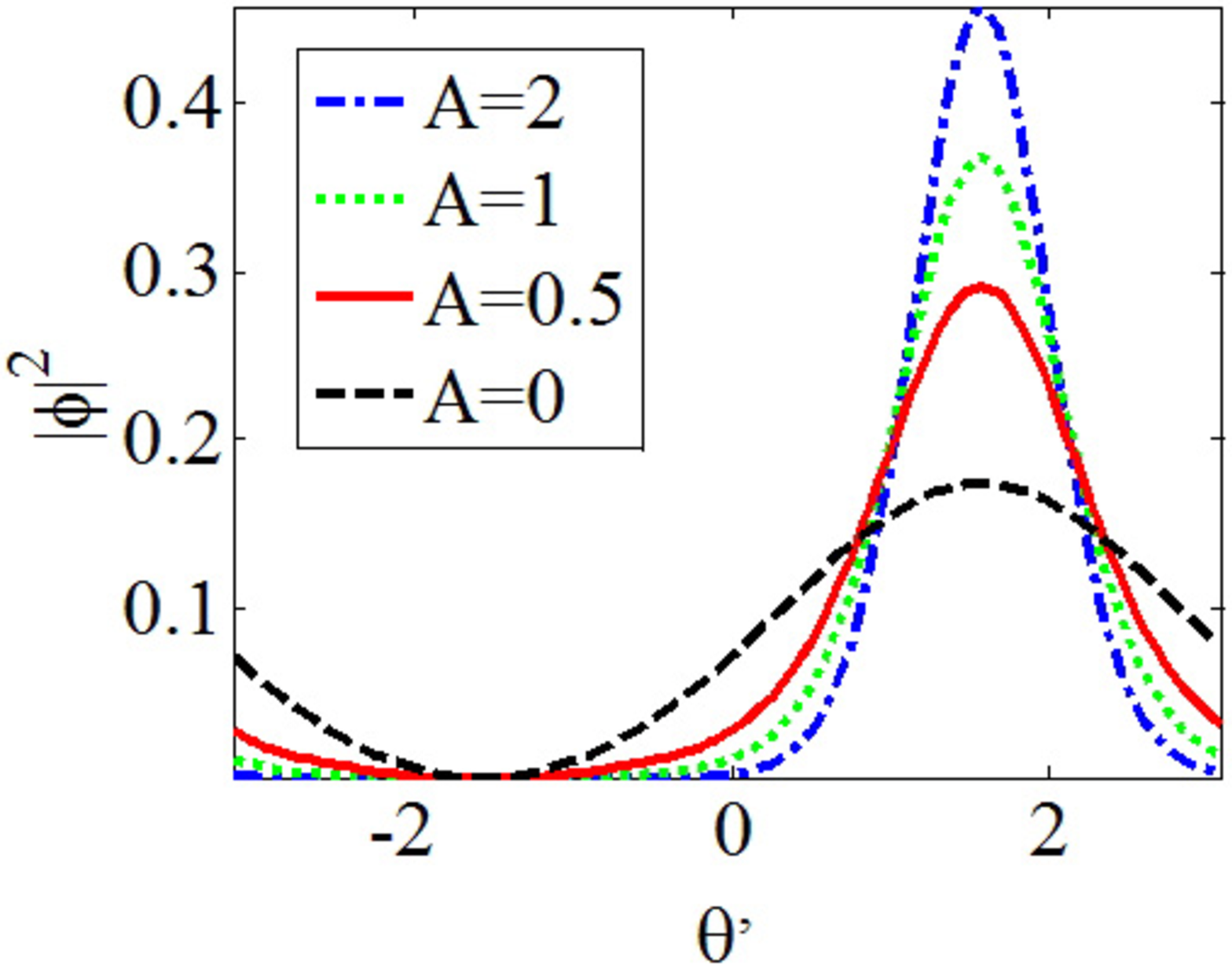}}
\subfigure[]{ \label{fig_5_c}
\includegraphics[scale=0.21]{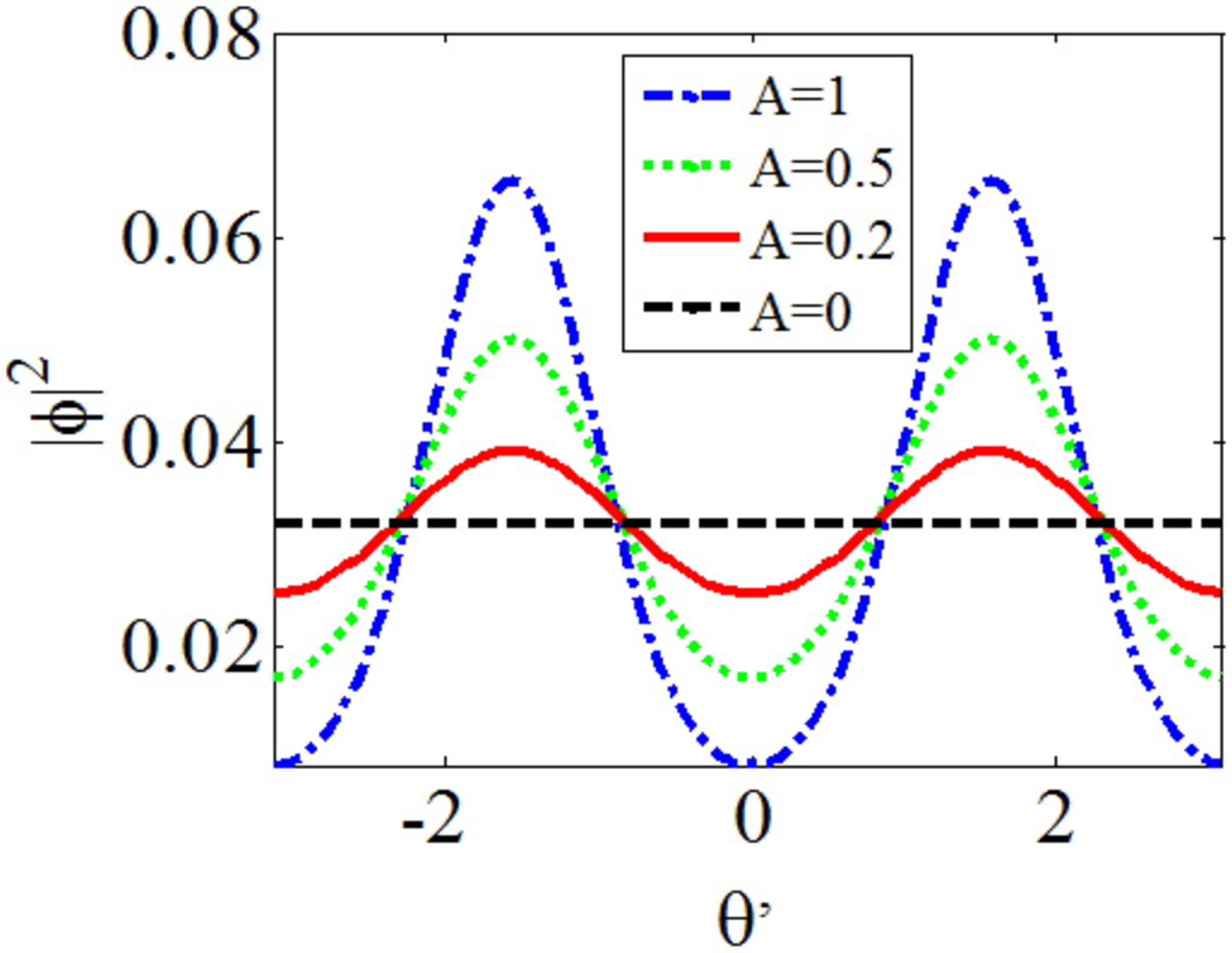}}
\subfigure[]{ \label{fig_5 d}
\includegraphics[scale=0.21]{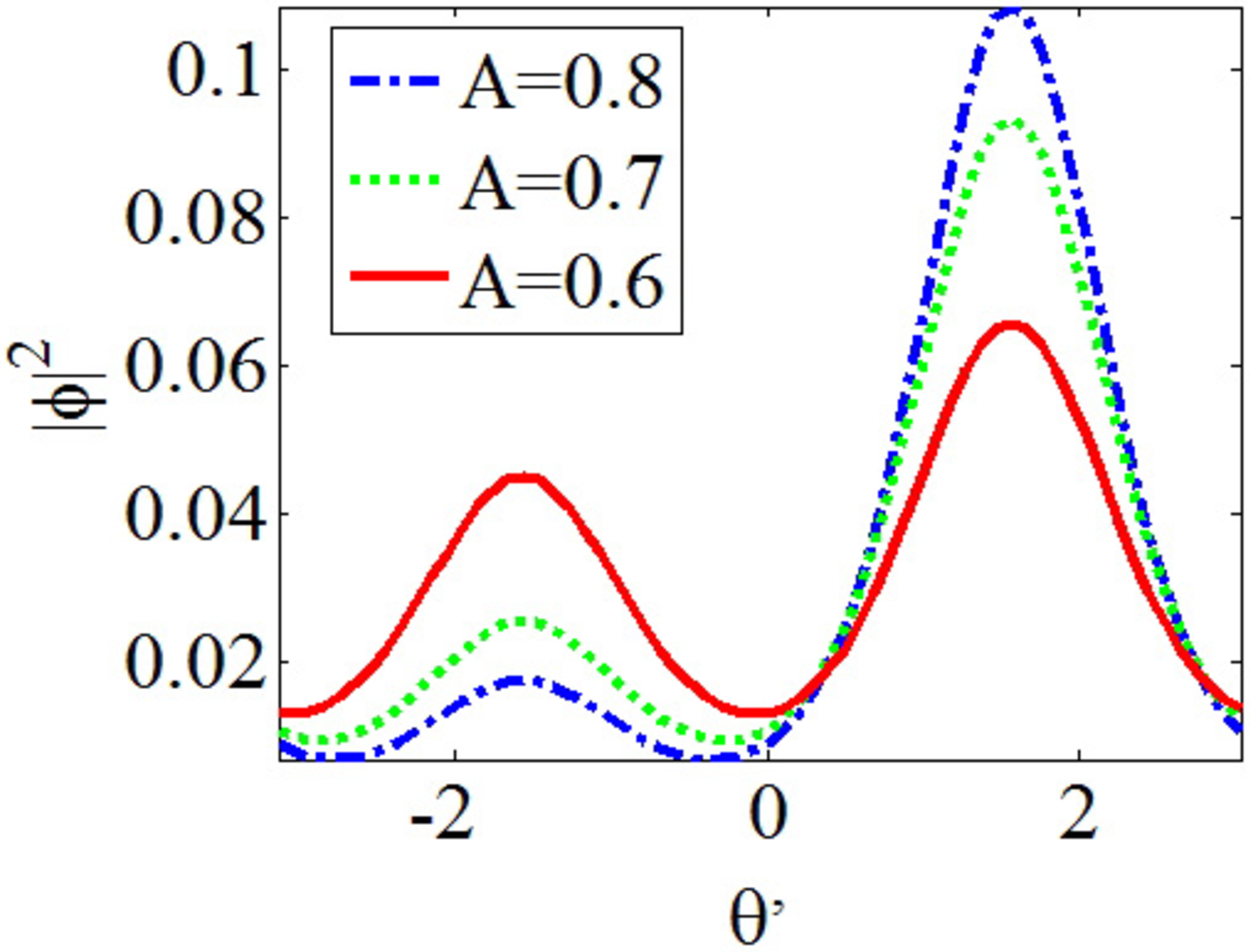}}
\caption{(Color online) (a) The FHS mode, in the case of the SF nonlinearity
with $(\protect\omega ,P)=(0.25,0.2)$, at different values of the
potential's strength, $A$. (b) The same for the FHA mode with $(\protect%
\omega ,P)=(0.5,0.5)$. (c) The same for the 2HS mode with $(\protect\omega %
,P)=(0.25,0.2)$. (d) The same for the 2HA mode with $(\protect\omega %
,P)=(0.4,0.2)$. All the modes shown in this figure are stable.}
\label{Amod}
\end{figure}

In the limit of the uniform ring (without the potential, $A=0$) the FHS and
AnS modes go over into the above-mentioned uniform vortical state (\ref{1}),
the 2HS pattern degenerates into the CW state (\ref{0}), while the FHA mode
takes the form of a $2\pi $-periodic cnoidal-wave solution of the nonlinear
Schr\"{o}dinger equation equation (the 2HA pattern does not exist at all in
the limit of $A=0$). The evolution of the shape of the modes with the
increase of $A$ is shown in Fig. \ref{Amod} [The evolution figure of AnS mode, which is not shown here, is similar to the 2HS type displayed in Fig. \ref{fig_5_c}].

\subsection{Existence and stability diagrams for the different modes}

Results of the systematic numerical analysis are summarized in Fig. \ref%
{1stPomega}, in the form of diagrams for the existence of the stable FHS and
FHA modes in the plane of $\left( P,\omega \right) $ at several fixed values
of amplitude $A$ of the harmonic potential. In the blank areas, only \emph{%
unstable} modes of the FHS type are found [in direct simulations they
feature an oscillatory instability, which is accounted for by a quartet of
complex eigenvalues generated by Eq. (\ref{lambda})], while the FHA modes
are completely stable in their existence region.

\begin{figure}[tbp]
\centering\subfigure[] {\label{fig_7_a}
\includegraphics[scale=0.2]{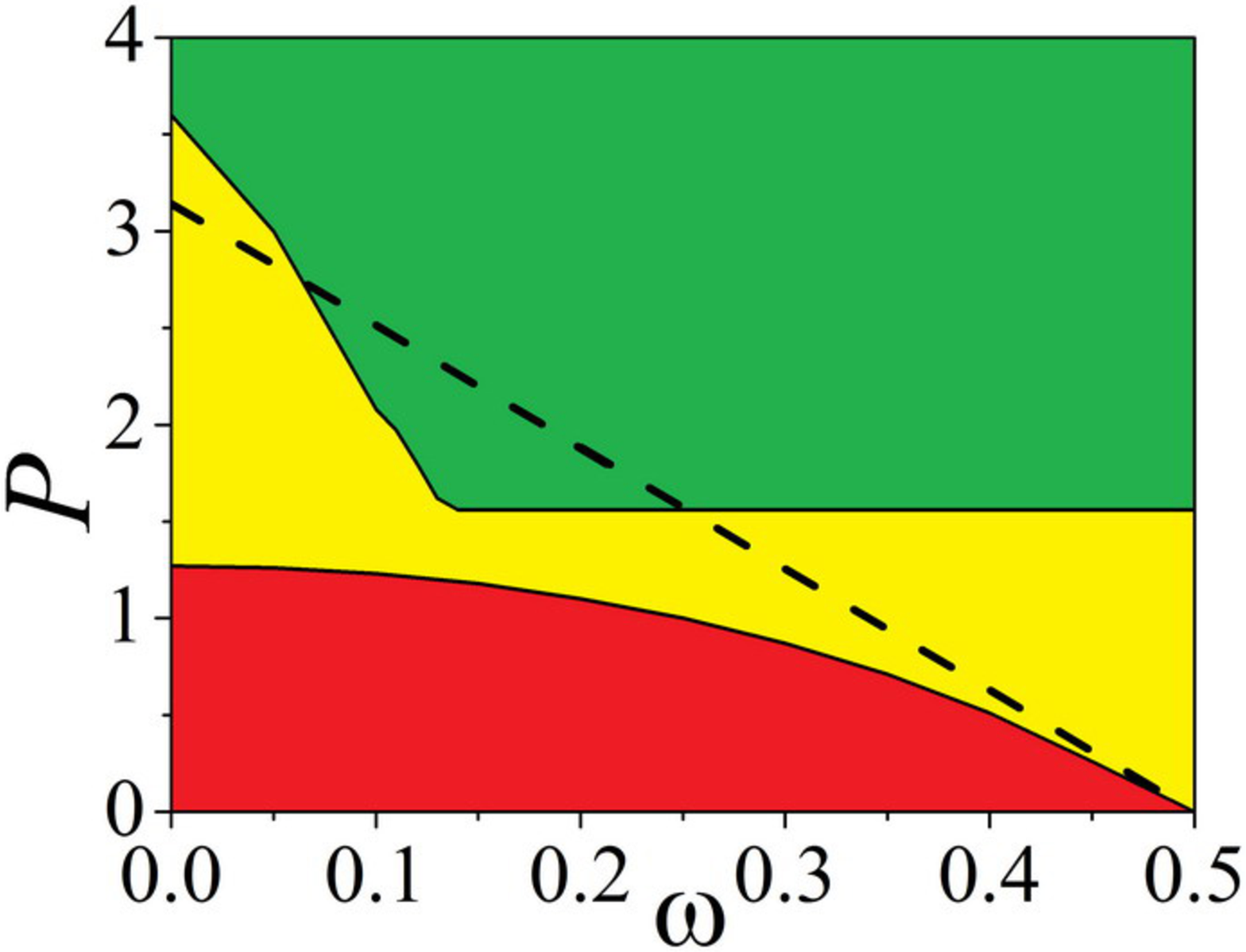}}%
\subfigure[] {\label{fig_7_b}
\includegraphics[scale=0.2]{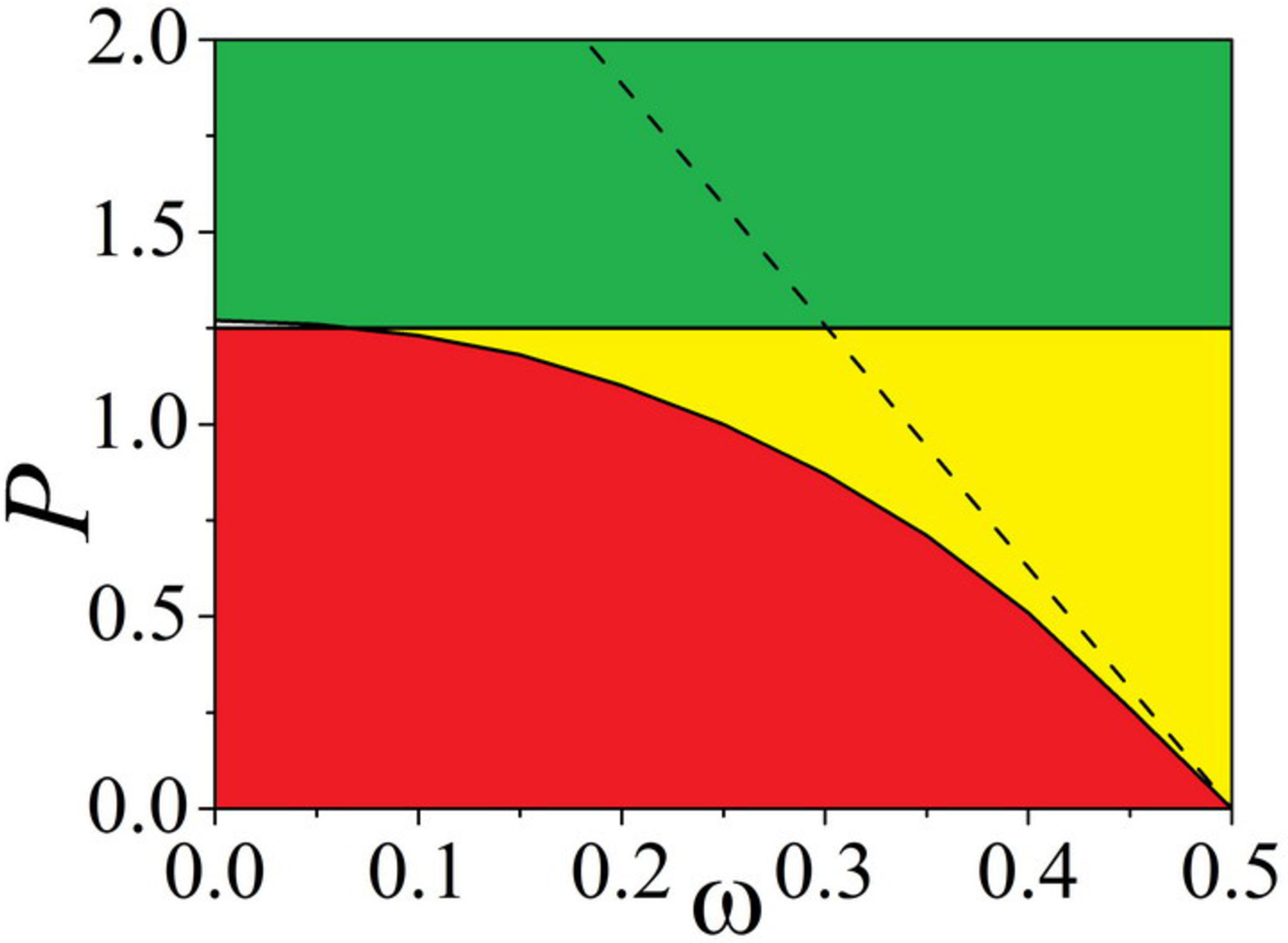}}
\subfigure[]{ \label{fig_7_c}
\includegraphics[scale=0.2]{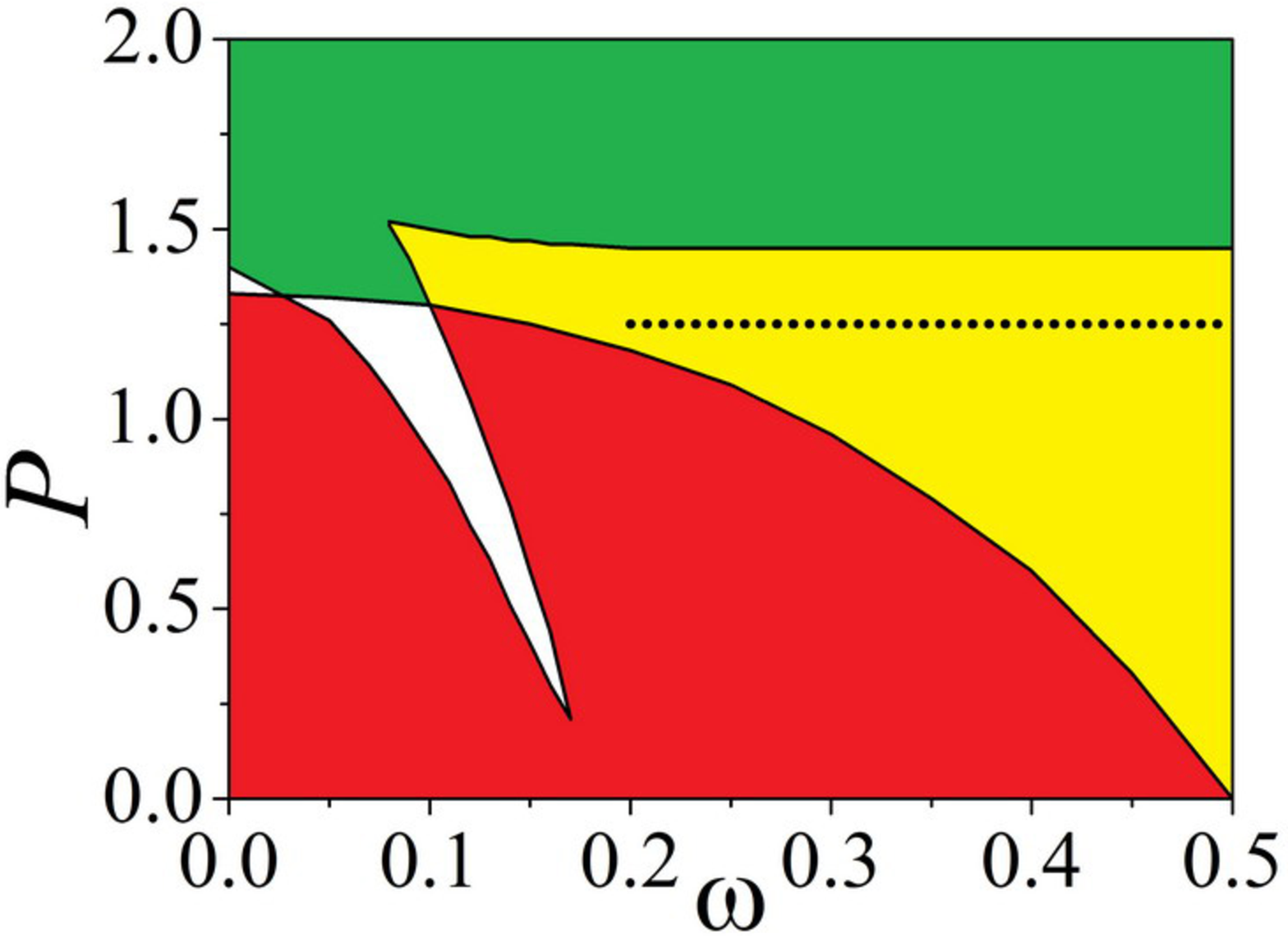}}
\subfigure[]{ \label{fig_7 d}
\includegraphics[scale=0.2]{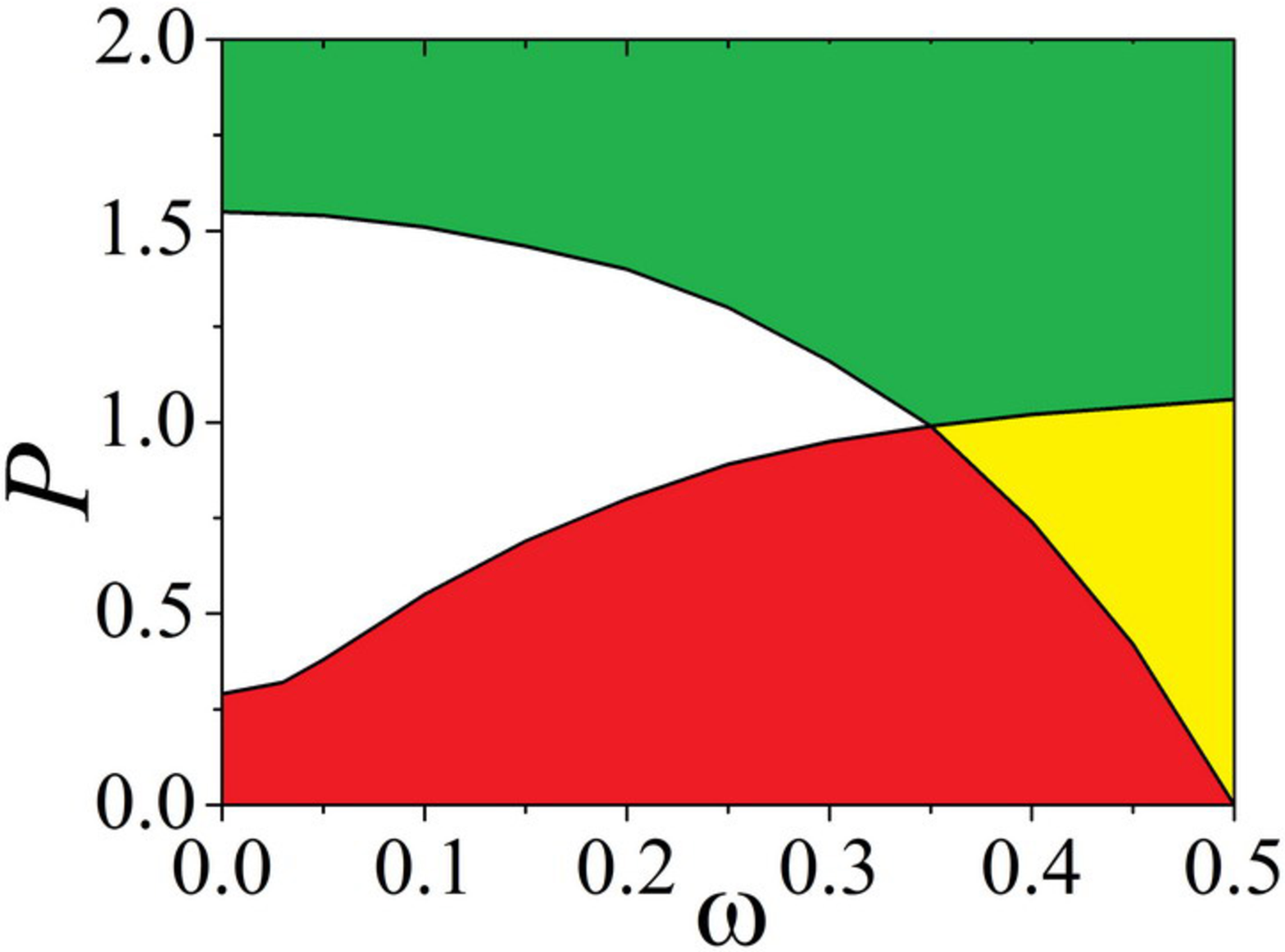}}
\caption{(Color online) Stability diagrams for the symmetric and asymmetric
modes dominated by the fundamental harmonic in the case of the SF
nonlinearity, in the plane of the rotation speed ($\protect\omega $) and
total power ($P$), at several fixed values of the potential's strength: (a) $%
A=0$, (b) $A=0.2$, (c) $A=0.5$, and (d) $A=1$. The green (top) and red
(bottom) areas designate regions where solely the FHS or FHA mode is stable,
respectively. The bistability (coexistence of these stable modes) occurs in
the yellow (middle) region. In the blank area, no stable modes of these
types are found (in fact, an\emph{\ }unstable FHS mode exists in this area).
In panels (a) and (b), the dashed lines represent the analytical prediction (%
\protect\ref{P}), $P=2\protect\pi |1/2-\protect\omega |$, for the existence
boundary of the FHA mode (the prediction is relevant at small values of $%
|1/2-\protect\omega |$, $P$, and $A$). The horizontal dotted line in (c) is
the cross section of the bistability region, along which energies of the FHS
and FHA modes is compared in Fig. \protect\ref{1stHam}(a).}
\label{1stPomega}
\end{figure}

The dashed lines in panels (a) and (b) of Fig. \ref{1stPomega} demonstrate
that Eq. (\ref{P}) correctly predicts the bistability boundary at small
values of $\delta \equiv 1/2-\omega $ and $P$, provided that $A$ is small
enough too [recall it is the condition under which Eq. (\ref{P}) was
derived]. Further, the fact that vertical cuts of the panels in Fig. \ref%
{1stPomega}, that can be drawn through $\omega =\mathrm{const}$, go, with
the increase of the total power ($P$), from the region of the monostability
of the FHS mode through the FHS/FHA bistability area into the monostability
region of the FHA mode (at least, for $\omega $ sufficiently close to $1/2$%
), clearly suggests that the symmetry-breaking bifurcation of the \textit{%
subcritical type}, which features the bistability \cite{Bif}, \cite{Snyder},
occurs along this route. This conclusion is confirmed by a direct
consideration of the numerical data (not shown here in detail).

\begin{figure}[tbp]
\centering%
\subfigure[] {\label{fig_11add_a}
\includegraphics[scale=0.2]{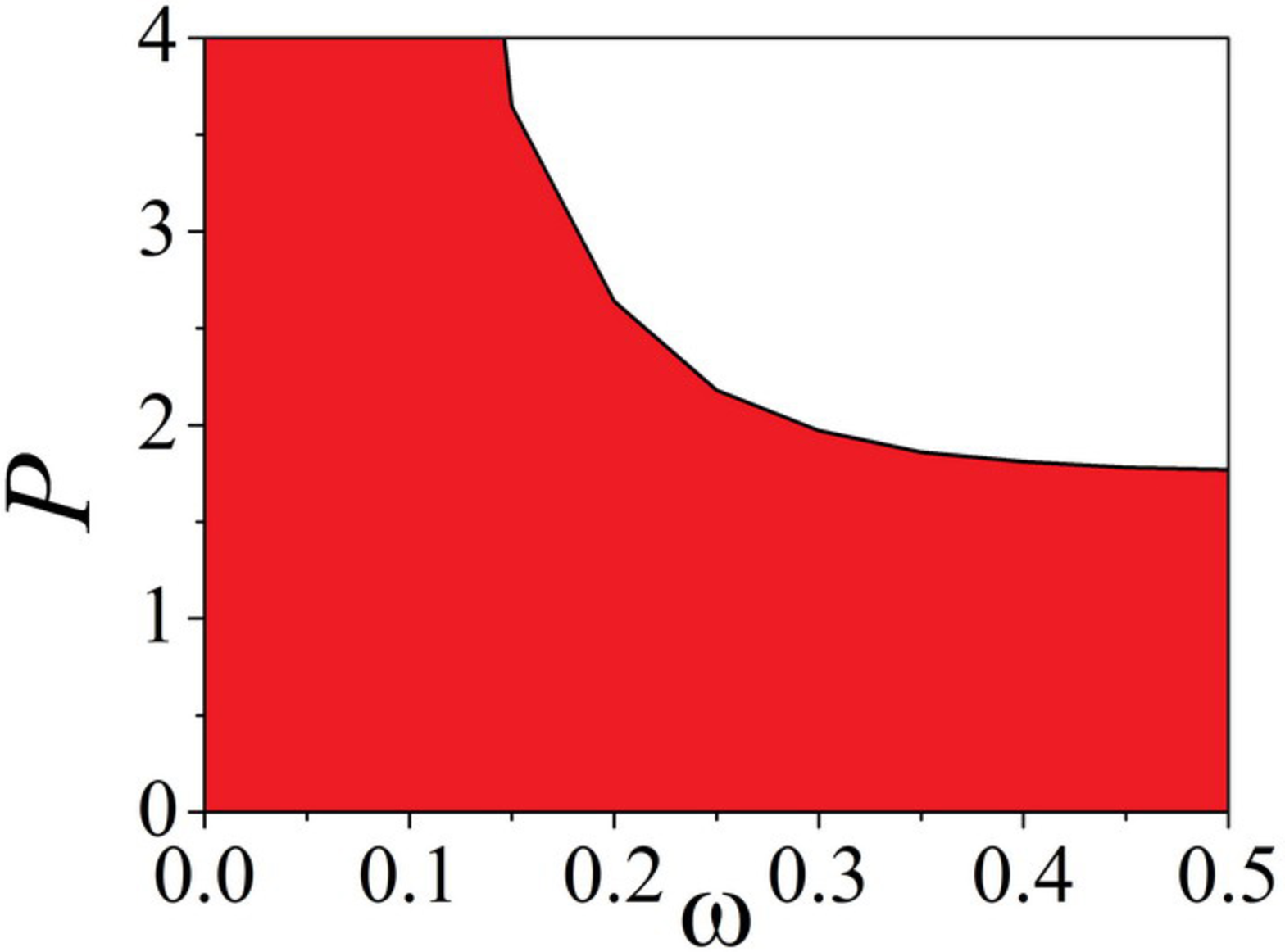}}%
\subfigure[] {\label{fig_11add_b}
\includegraphics[scale=0.2]{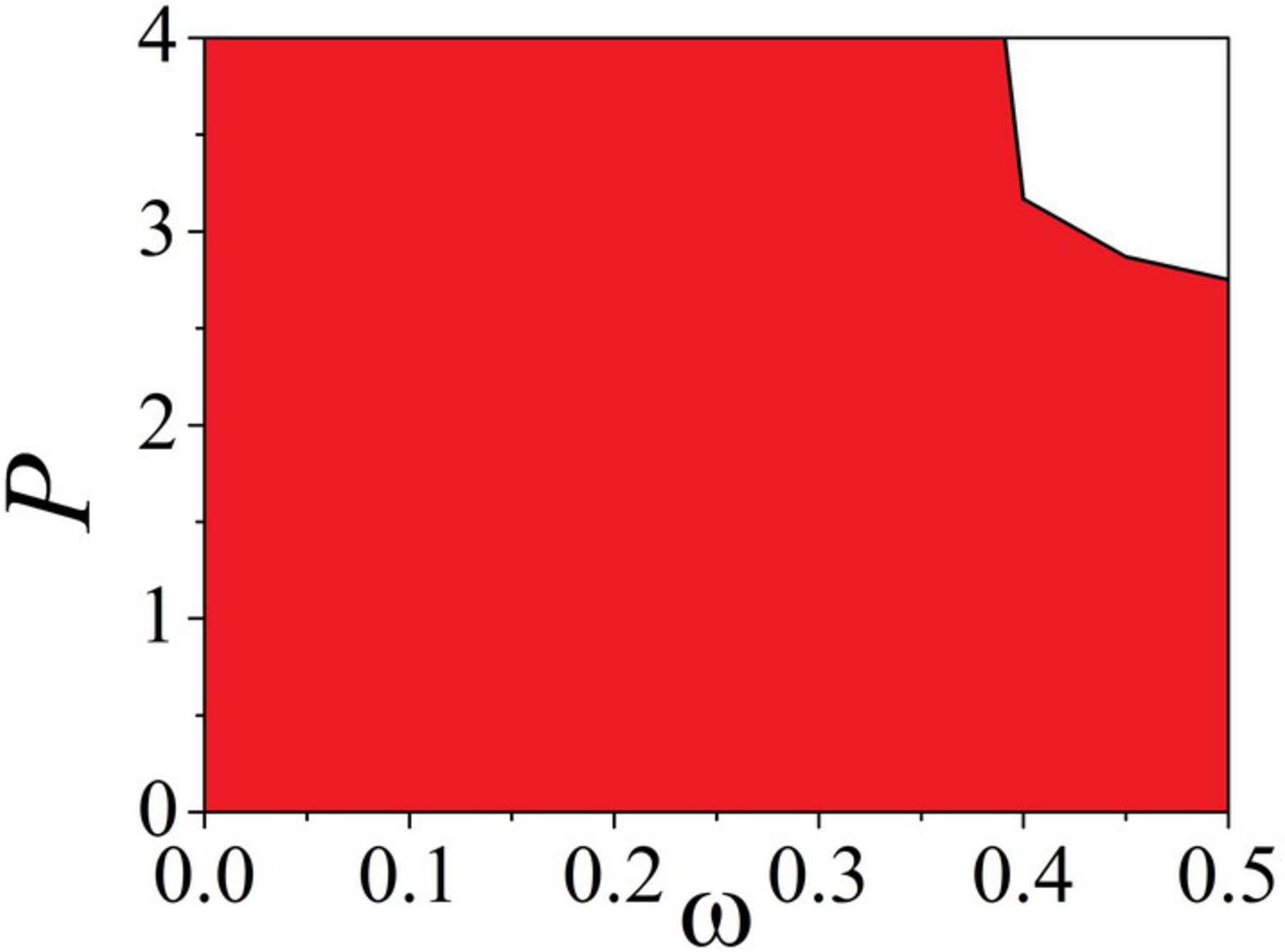}}
\caption{(Color online) In the case of the SF nonlinearity, the AnS
(antisymmetric) mode exists and is stable in the red area, as shown here for
(a) $A=0.2$ and $A=0.4$ (b). }
\label{antiSFPomega}
\end{figure}

As concerns the AnS solitons, in the case of the SF nonlinearity they do not
undergo any bifurcation, and are stable in the entire region of their
existence, which is shown in the $(\omega ,P)$ plane for fixed values of the
lattice's strength, $A$, in Fig. \ref{antiSFPomega} [as mentioned above, at $%
A=0$ the AnS mode coincides with the FHS one, hence the respective existence
area is the same as the red (bottom) one in Fig. \ref{1stPomega}(a)]. From
here we conclude that the stability region of the AnS mode quickly expands
with the increase of $A$, and the FHS/FHA bistability areas in panels (b)
and (c) of Fig. \ref{1stPomega} actually support the \textit{tristability},
as the AnS mode is stable there too.

\begin{figure}[tbp]
\centering%
\subfigure[] {\label{fig_9_a}
\includegraphics[scale=0.18]{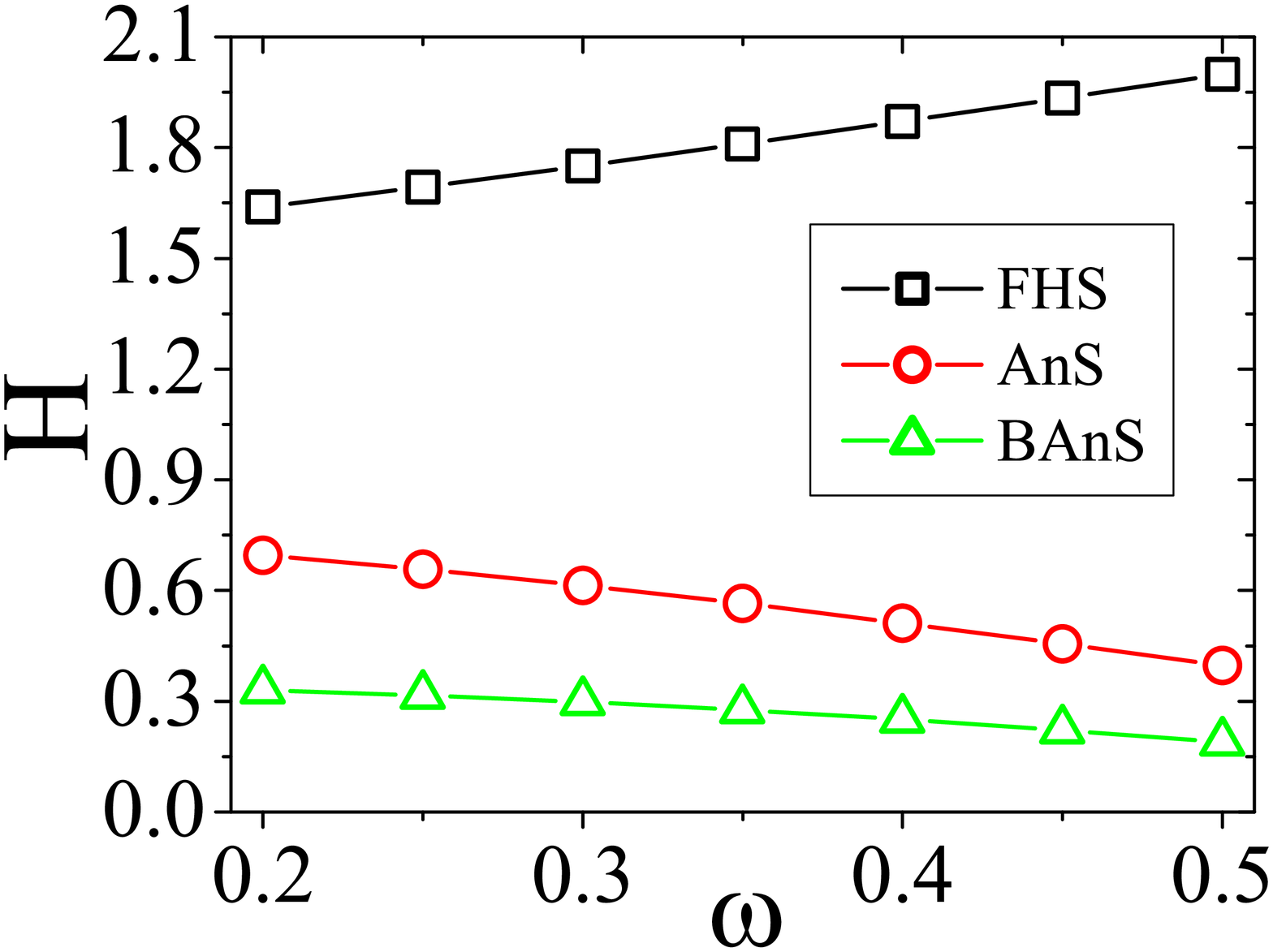}}%
\subfigure[] {\label{fig_9_b}
\includegraphics[scale=0.18]{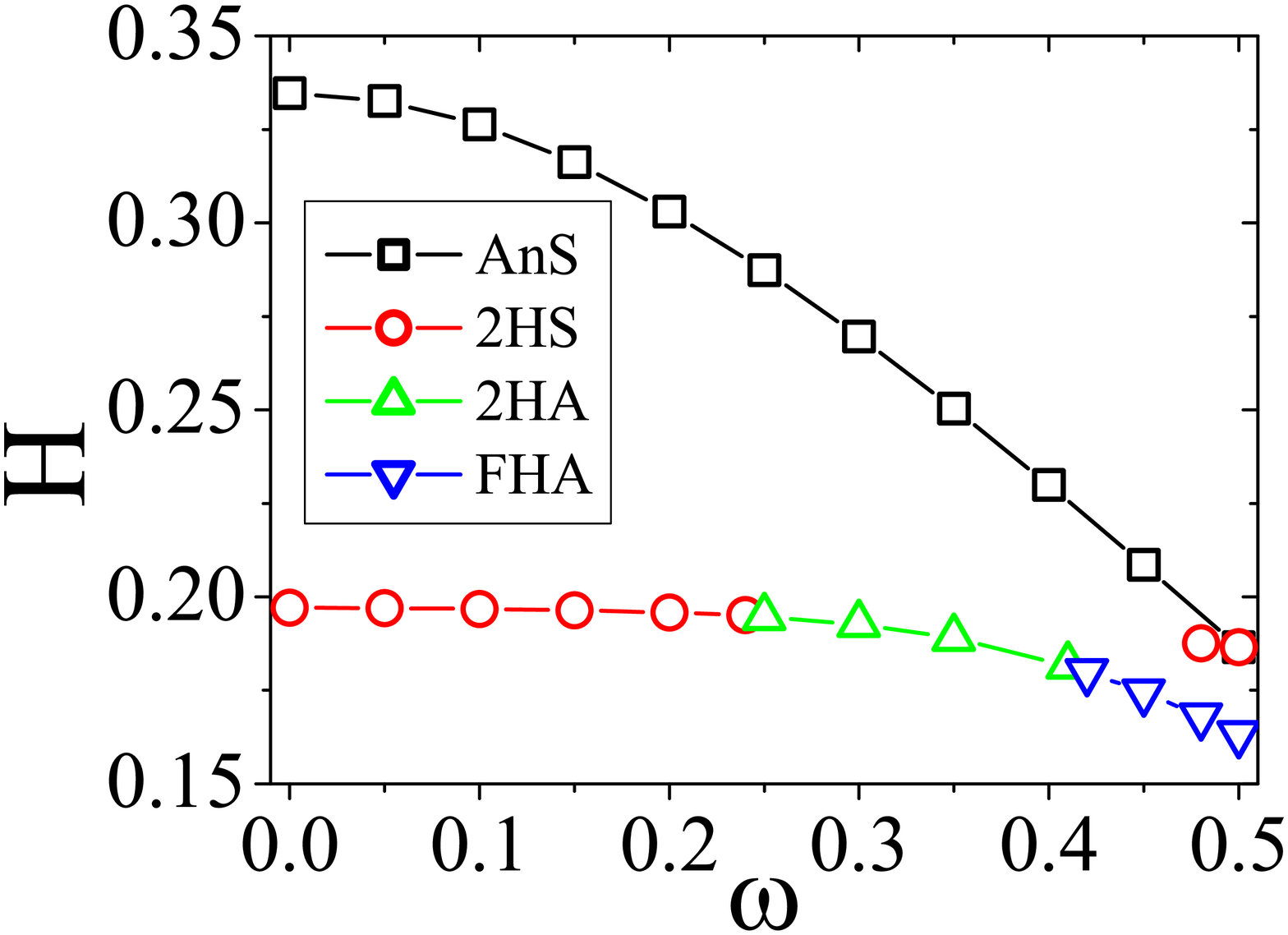}}
\subfigure[] {\label{fig_9_c}
\includegraphics[scale=0.18]{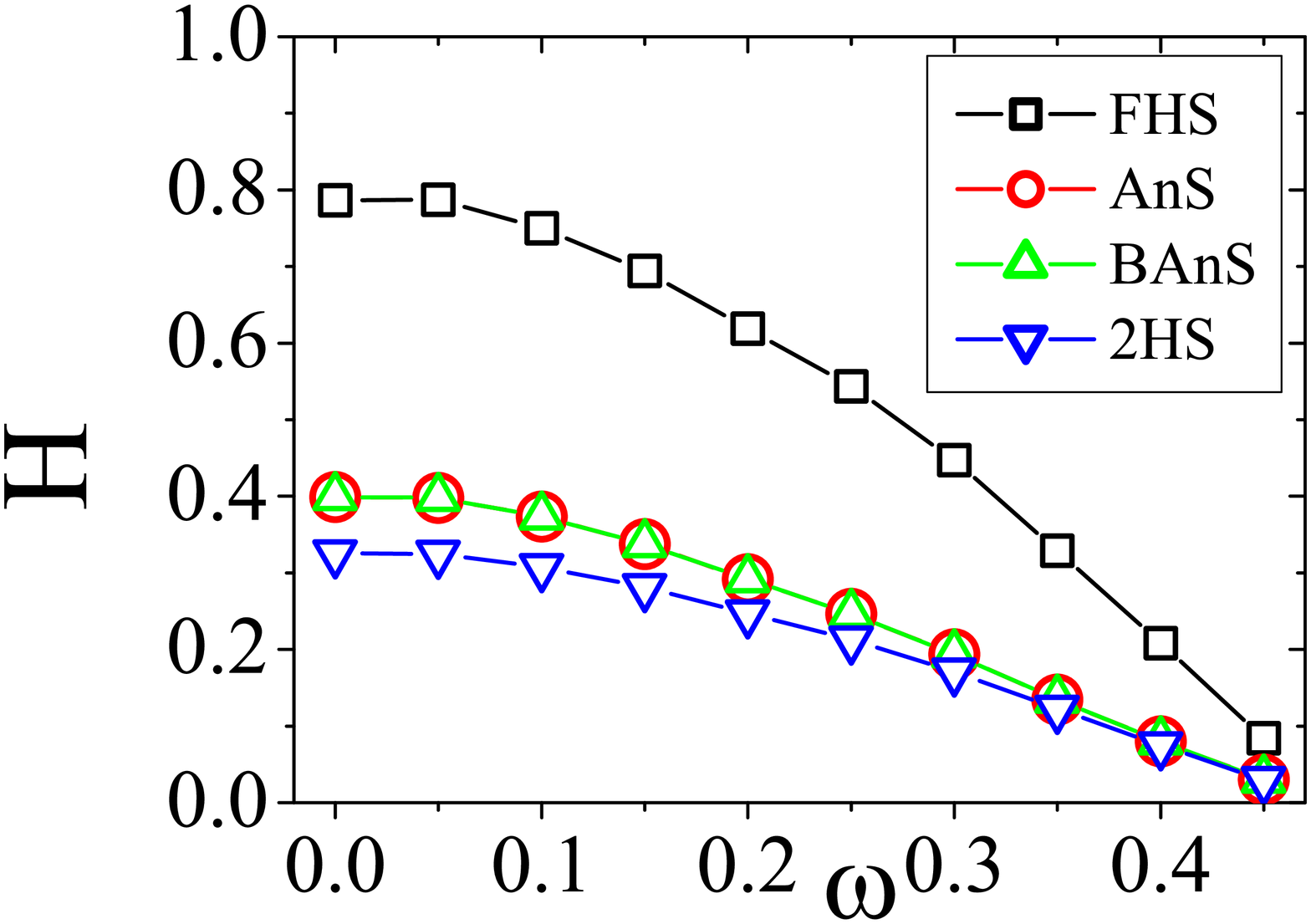}} \label{1stHam}
\caption{(Color online) (a) Energies of the FHS, FHA, and AnS modes,
computed according to Eq. (\protect\ref{Ham}) along the horizontal dotted
line drawn in Fig. \protect\ref{fig_7_c}, with $(P,A)=(1.25,0.5)$, in the
interval of $\ 0.2\leq \protect\omega \leq 0.5$. (b) The energies of the
2HS, 2HA, and FHA modes, along with the coexisting AnS one, along the
horizontal dotted line drawn in Fig. \protect\ref{fig_11_c}, for $%
(P,A)=(0.5,0.5)$. Panels (a) and (b) pertain to the system with the SF
nonlinearity, $\protect\sigma =+1$. (c) The energies of the AnS, BAnS
(broken-antisymmetry), and 2HS modes, along the boundary separating the
existence regions of the AnS and BAnS solutions in the case of the SDF
nonlinearity ($\protect\sigma =-1$), see Fig. \protect\ref{fig_15_b}.}
\end{figure}

The multistability between the three species of the modes based on the
fundamental harmonic---symmetric, asymmetric, and antisymmetric ones---makes
it necessary to compare their energies, which are defined as per Eq. (\ref%
{Ham}). The comparison along the horizontal dotted line, which is drawn in
Fig. \ref{fig_7_c}, is presented in\ Fig. \ref{fig_9_a}. The figure clearly
shows the following relation between the energies:
\begin{equation}
H_{\mathrm{FHA}}<H_{\mathrm{AnS}}<H_{\mathrm{FHS}}.  \label{HHH}
\end{equation}%

\begin{figure}[tbp]
\centering%
\subfigure[] {\label{fig_11_a}
\includegraphics[scale=0.2]{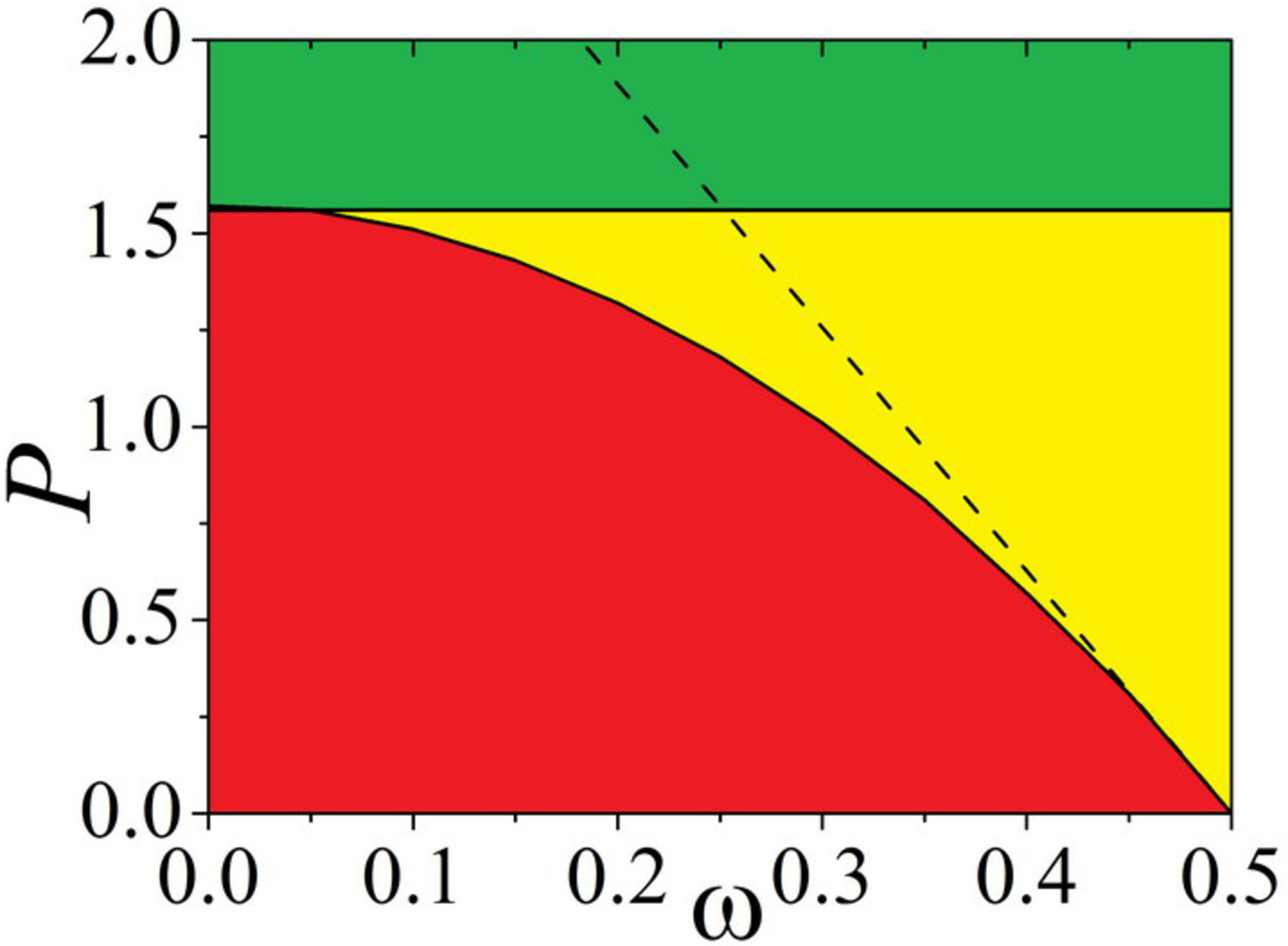}}%
\subfigure[] {\label{fig_11_b}
\includegraphics[scale=0.2]{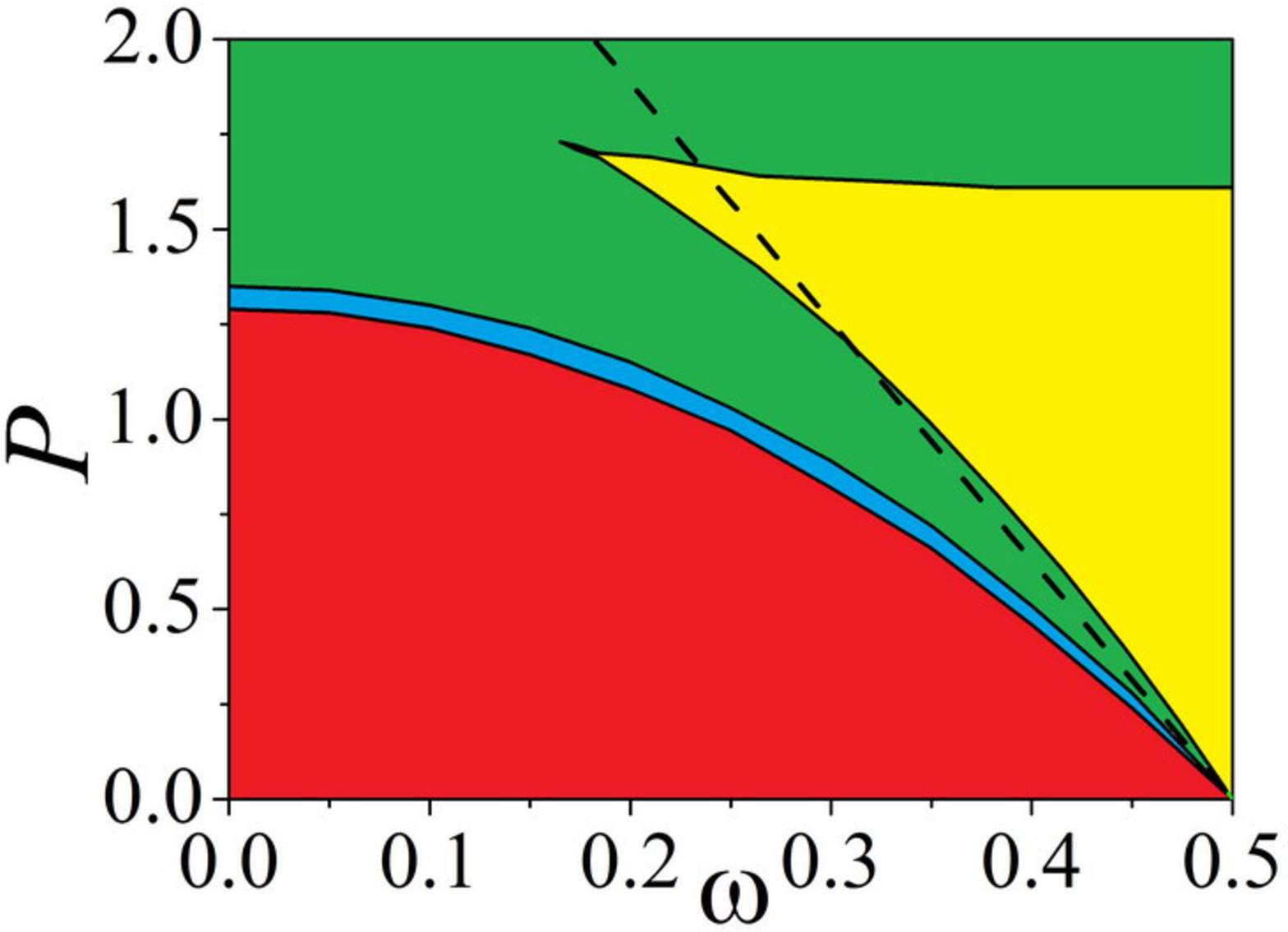}}
\subfigure[]{ \label{fig_11_c}
\includegraphics[scale=0.2]{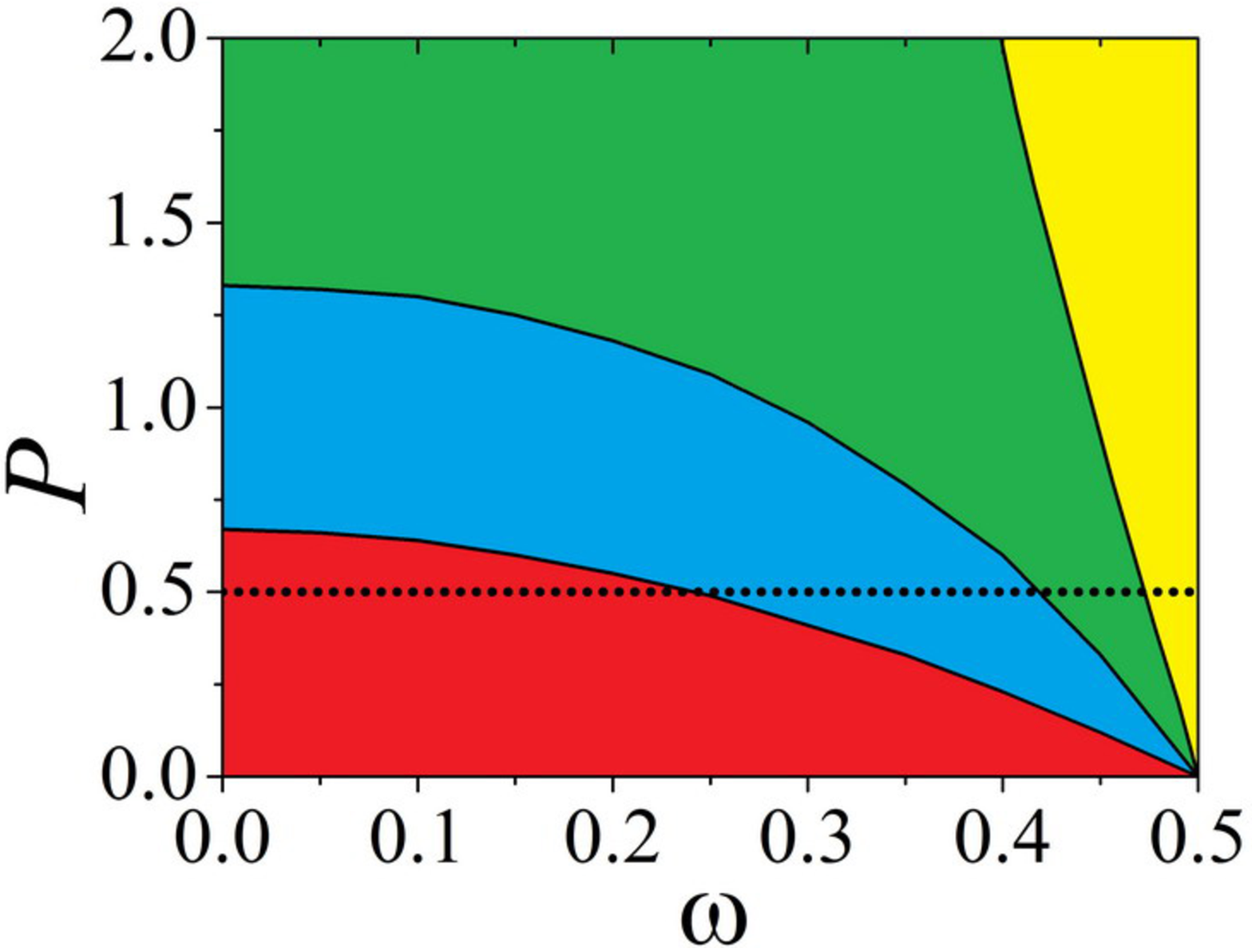}}
\subfigure[]{ \label{fig_11 d}
\includegraphics[scale=0.2]{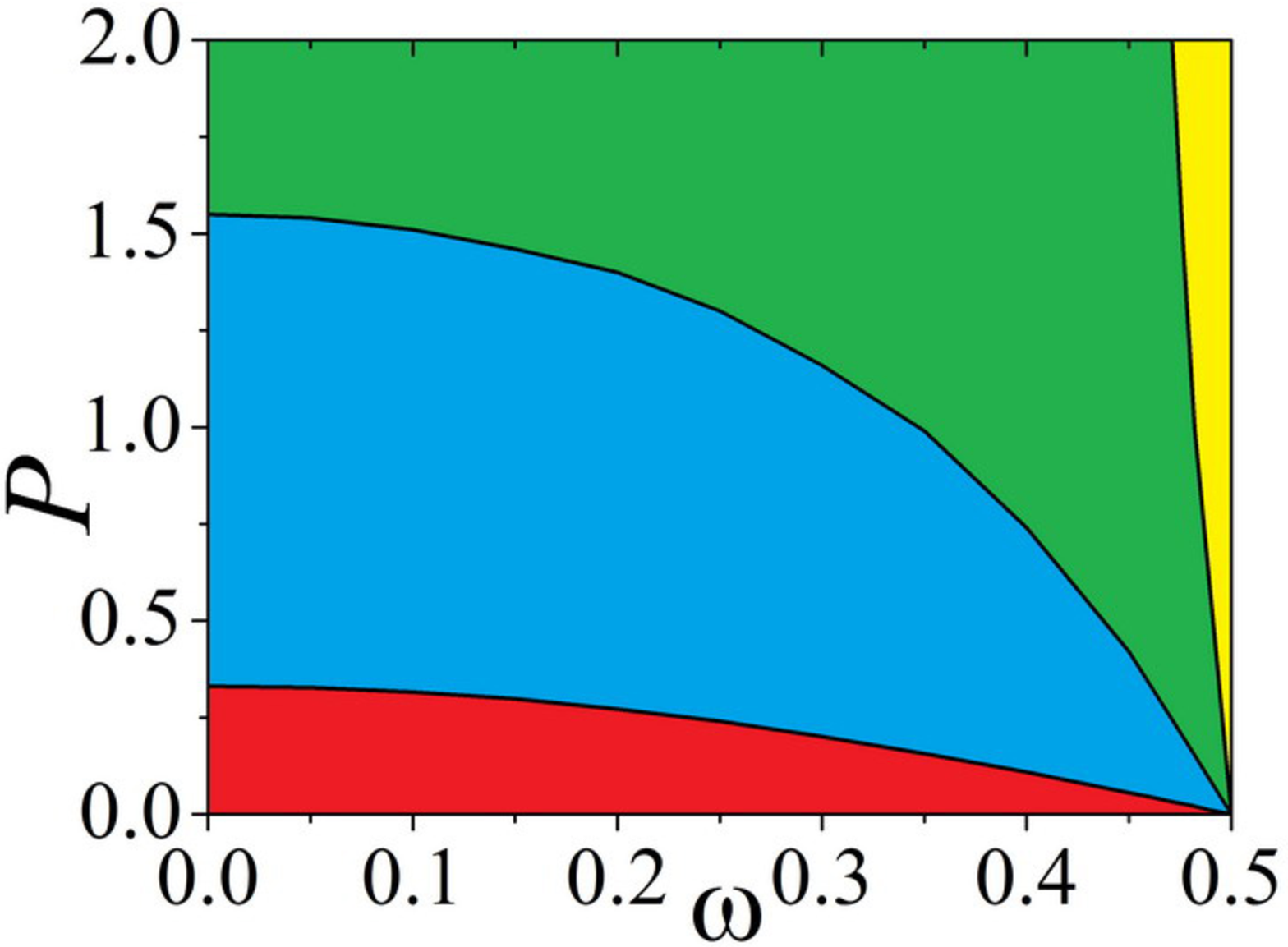}}
\caption{(Color online) The same as in Fig. \protect\ref{1stPomega}, but for
the following types of the solutions. The green (top) area: the stable FHA
mode; the red (bottom) area: the stable 2HS mode; the blue (intermediate)
area: the stable 2HA mode; the yellow (rightmost) area: the bistability of
the FHA and 2HS modes.}
\label{2ndPomega}
\end{figure}

For the 2HS and 2HA patterns, which are based on the second angular
harmonic, Fig. \ref{2ndPomega} displays the stability areas in the $\left(
P,\omega \right) $ plane, at different values of the lattice's strength, $A$
[cf. Fig. \ref{1stPomega}], together with the stability area for the FHA
mode, which may stably coexist with the 2HS pattern. Figure \ref{2ndPomega}
also demonstrates that, as mentioned above, the 2HA mode, which exists and
is stable in the blue (intermediate) area in Fig. \ref{2ndPomega}, does not
exist at $A=0$, emerging and expanding with the increase of $A$. Note that,
on the contrary to the situation for the patterns dominated by the
fundamental angular harmonic (FHS and FHA) described above, there is no
overlap between the stability areas of the symmetric and asymmetric
second-harmonic-based modes, 2HS and 2HA, hence (as confirmed by an
additional consideration of the numerical results) the symmetry-breaking 2HS
$\rightarrow $ 2HA transition, following the increase of $P$ at $A\neq 0$,
amounts to a \textit{supercritical} bifurcation, which features no
bistability \cite{Bif}, \cite{Snyder}.

Another peculiarity of the symmetric mode based on the second harmonic (2HS)
is that it is stable in two \emph{disjoint} stability areas---the red and
yellow ones in Fig. \ref{2ndPomega} (the bottom and rightmost regions), in
the latter one the 2HS featuring the bistability with the FHA mode, although
they are not linked by any bifurcation. On the other hand, there is no
overlap between the stability areas of the asymmetric modes based on the
different angular harmonics, FHA and 2HA, therefore these modes may be
generated by the same input waveform ($1+\sin \theta ^{\prime }$, in Table
1), in different parameter regions. 

The effective bistability occurring in Fig. \ref{2ndPomega} makes it
necessary to compare the energy of the coexisting stable patterns, which is
done in Fig. \ref{fig_9_b}, along the horizontal dotted line drawn in Fig. %
\ref{2ndPomega}(c). The energy of the coexisting stable AnS mode is included
too. In particular, the short segment corresponding to the 2HS mode appears,
above the one pertaining to the FHA mode, when the horizontal dotted line in
Fig. \ref{2ndPomega}(c) enters the yellow (rightmost) region of the 2HS-FHA
bistability. Thus, adding the results from Fig. \ref{fig_9_b}, we extend
energy relations (\ref{HHH}) as follows:%
\begin{equation}
H_{\mathrm{FHA}}~\mathrm{or}~H_{\mathrm{2HA}}<H_{\mathrm{2HS}}<H_{\mathrm{AnS%
}}<H_{\mathrm{FHS}}.  \label{HHHHH}
\end{equation}%
The conclusion is that the asymmetric modes, either FHA or 2HA, represent
the \textit{ground state} of the rotating ring carrying the harmonic
potential and SF nonlinearity.

\section{Numerical results for the \textbf{self-defocusing nonlinearity (}$%
\protect\sigma =-1$\textbf{)}}

\begin{figure}[tbp]
\centering%
\subfigure[] {\label{fig_13_a}
\includegraphics[scale=0.2]{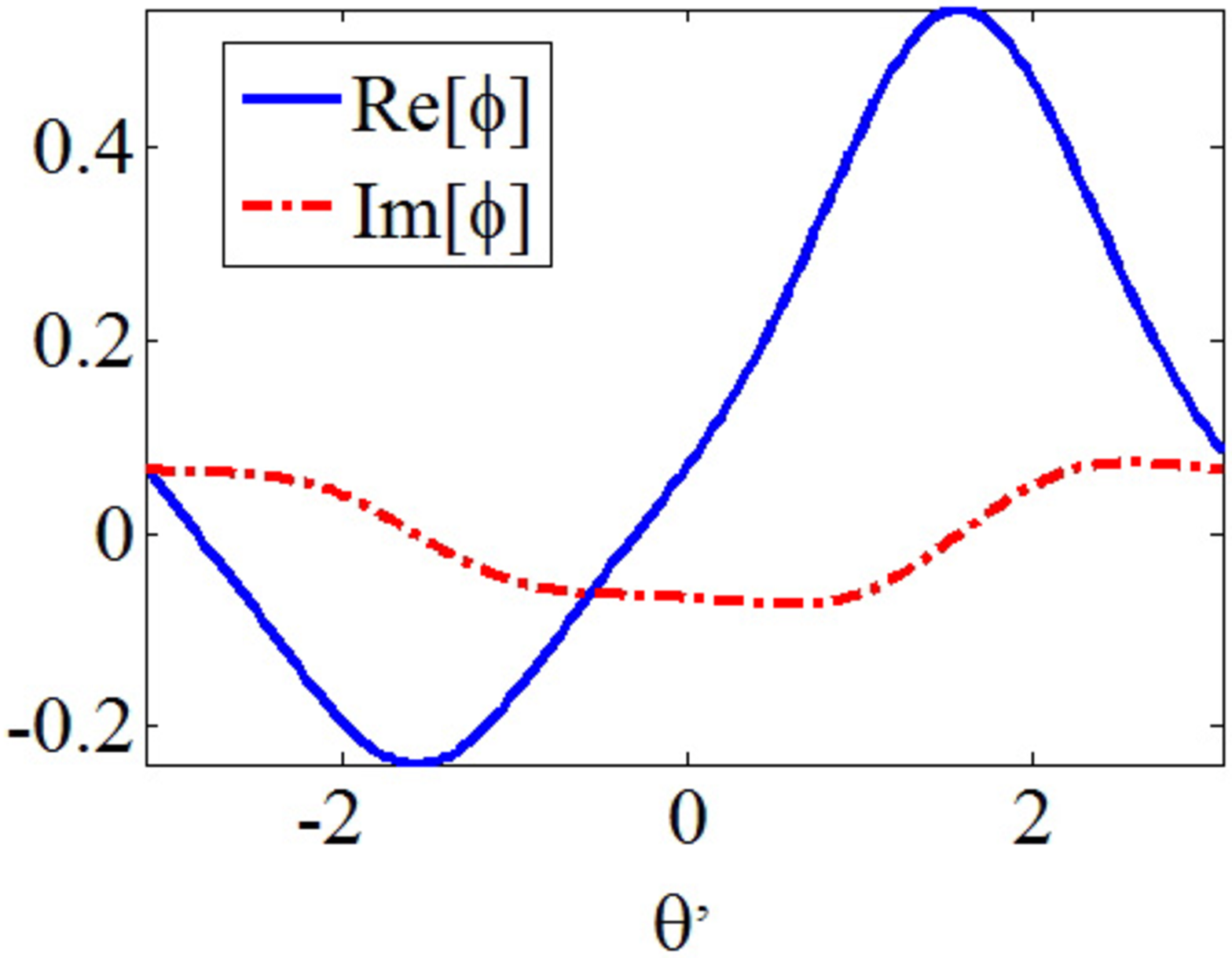}}%
\subfigure[] {\label{fig_13_b}
\includegraphics[scale=0.2]{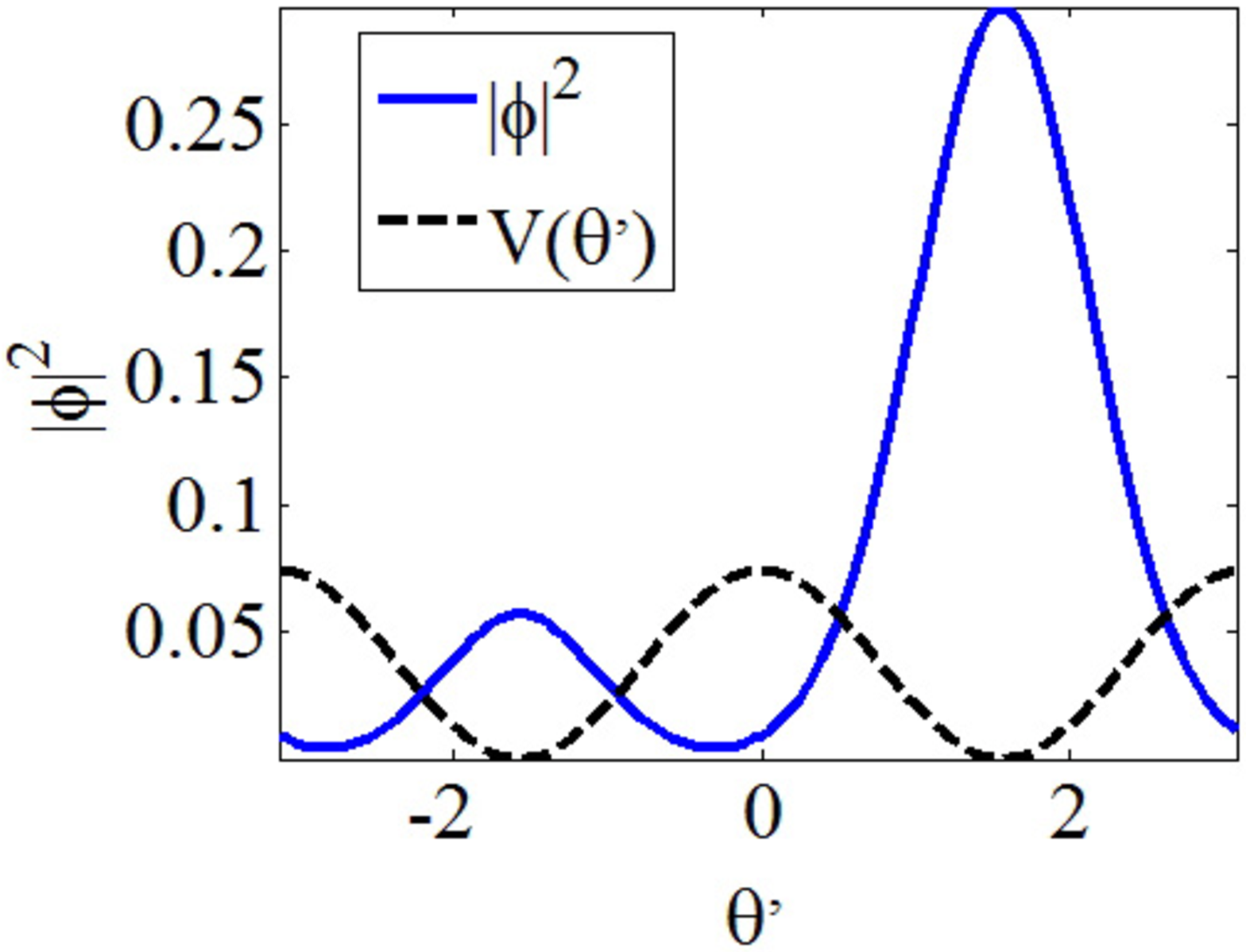}}
\caption{(Color online) A typical example of a stable BAnS mode (the one
with the \textit{broken antisymmetry}) in the system with the SDF
(self-defocusing) nonlinearity, $\protect\sigma =-1$, and parameters $(%
\protect\omega ,A,P)=(0.2,1,0.5)$: (a) the real and imaginary parts of the
stationary solution; (b) its intensity profile, which features unequal peaks
in the two potential wells.}
\label{Antiasy}
\end{figure}

In the case of the SDF nonlinearity, symmetric modes of the FHS and 2HS
types have been found too, although, on the contrary to the system with the
SF sign, they do not undergo any bifurcations, hence the FHA and 2HA species
do not exist in this case. On the other hand, the AnS mode features an
\textit{antisymmetry-breaking} bifurcation, which gives rise to a \textit{%
broken-antisymmetry} (BAnS) mode (which does not exist under the SF
nonlinearity). A typical example of the stable BAnS mode is displayed in
Fig. \ref{Antiasy}. In fact, antisymmetry-breaking bifurcations are
characteristic to the double-well systems with the SDF nonlinearity \cite%
{Warsaw}. The stationary solutions of the BAnS type can be numerically
generated by the initial guess of the form of $(b+\sin \theta ^{\prime })$,
with $0<b\leq 1$, see Table 1.

\begin{figure}[tbp]
\centering%
\subfigure[] {\label{fig_14_a}
\includegraphics[scale=0.2]{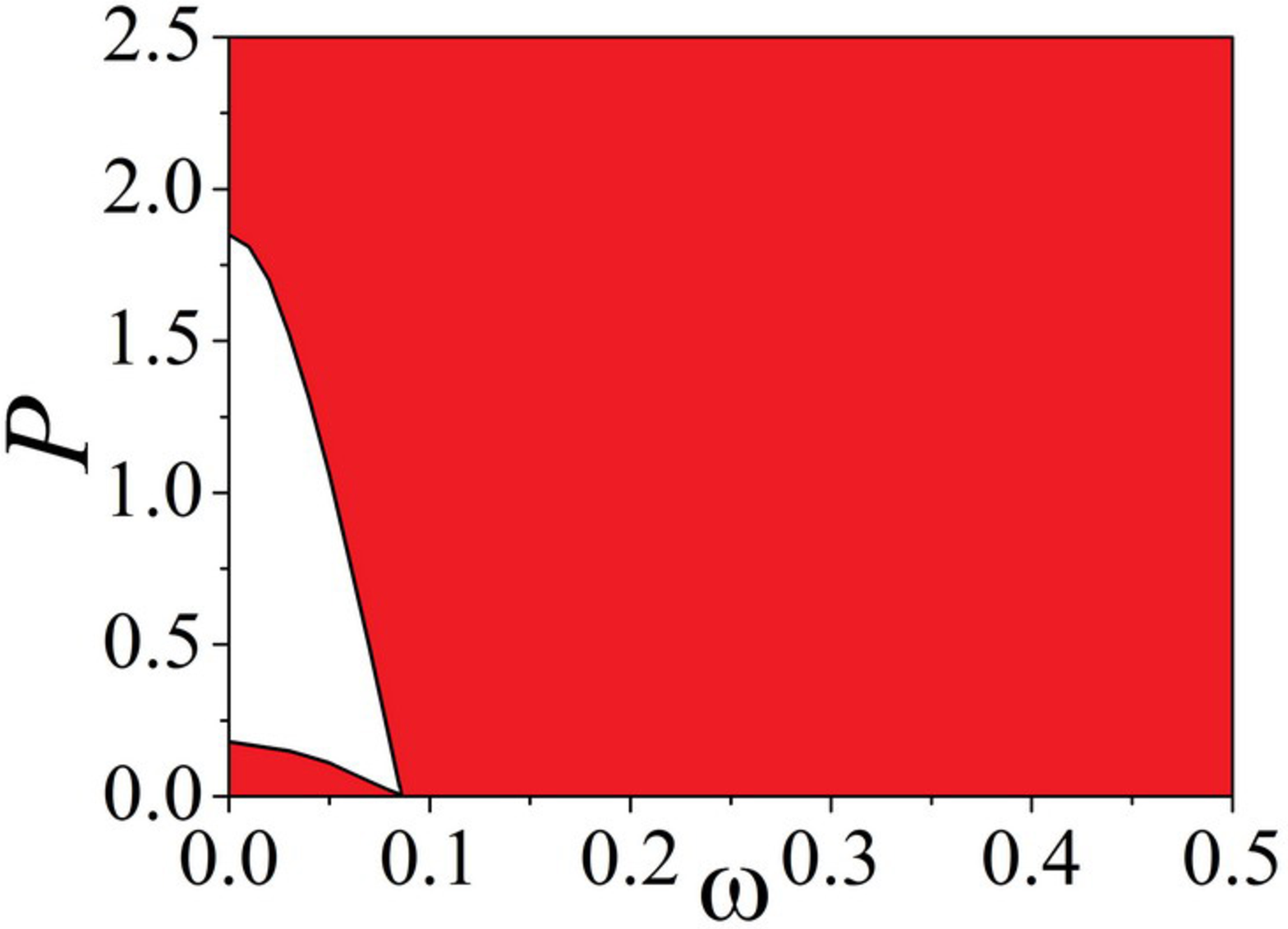}}%
\subfigure[] {\label{fig_14_b}
\includegraphics[scale=0.2]{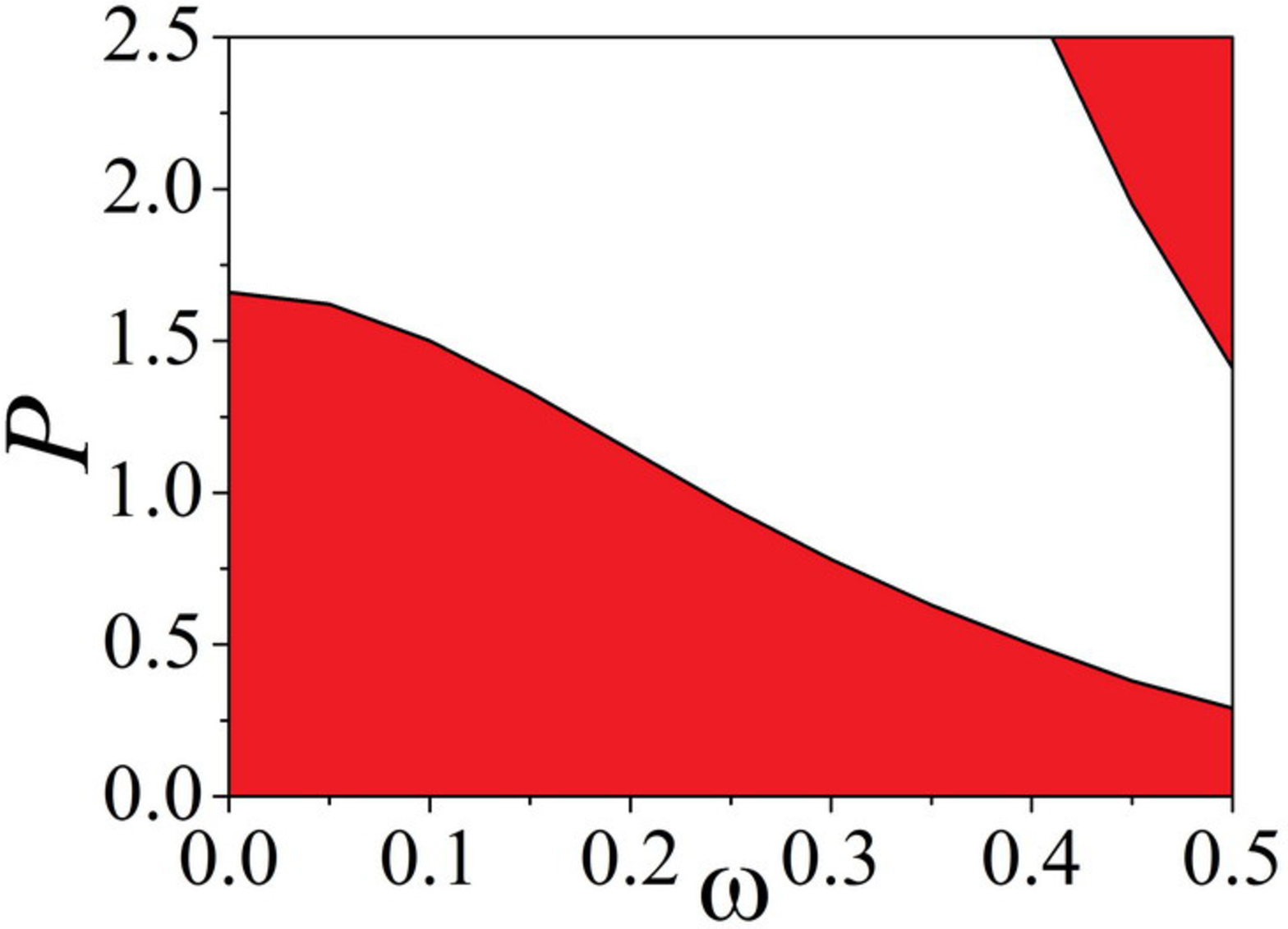}} 
\caption{(Color online) The stability diagram for the FHS mode, in the case
of the SDF nonlinearity ($\protect\sigma =-1$), in the $\left( \protect%
\omega ,P\right) $ plane, for $A=1.2$ (a) and $A=1.75$ (b). The FHS modes
are stable and unstable in the red and blank areas, respectively.}
\label{AntiPAomegaSDF}
\end{figure}

\begin{figure}[tbp]
\centering%
\subfigure[] {\label{fig_15_a}
\includegraphics[scale=0.2]{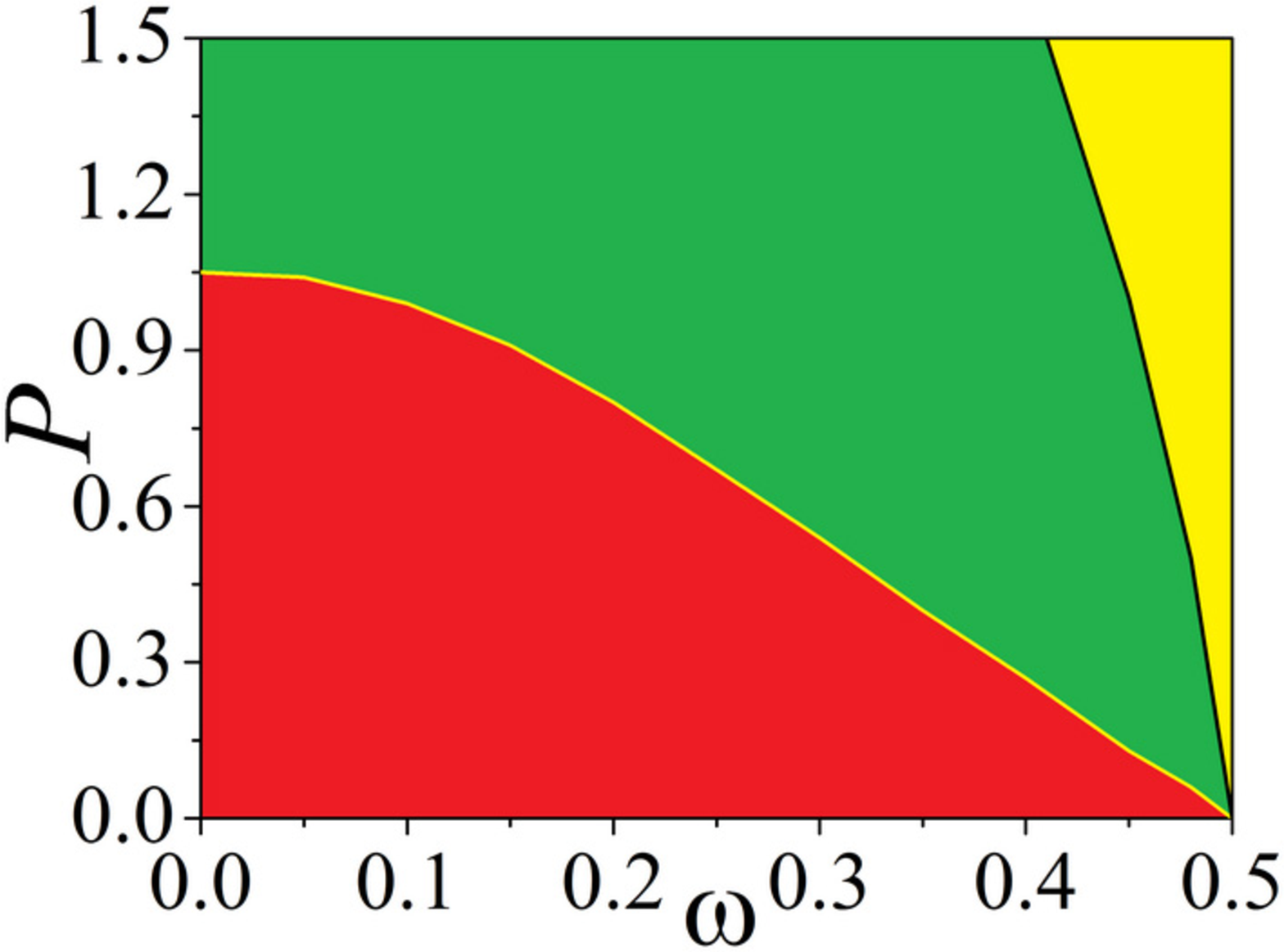}}%
\subfigure[] {\label{fig_15_b}
\includegraphics[scale=0.2]{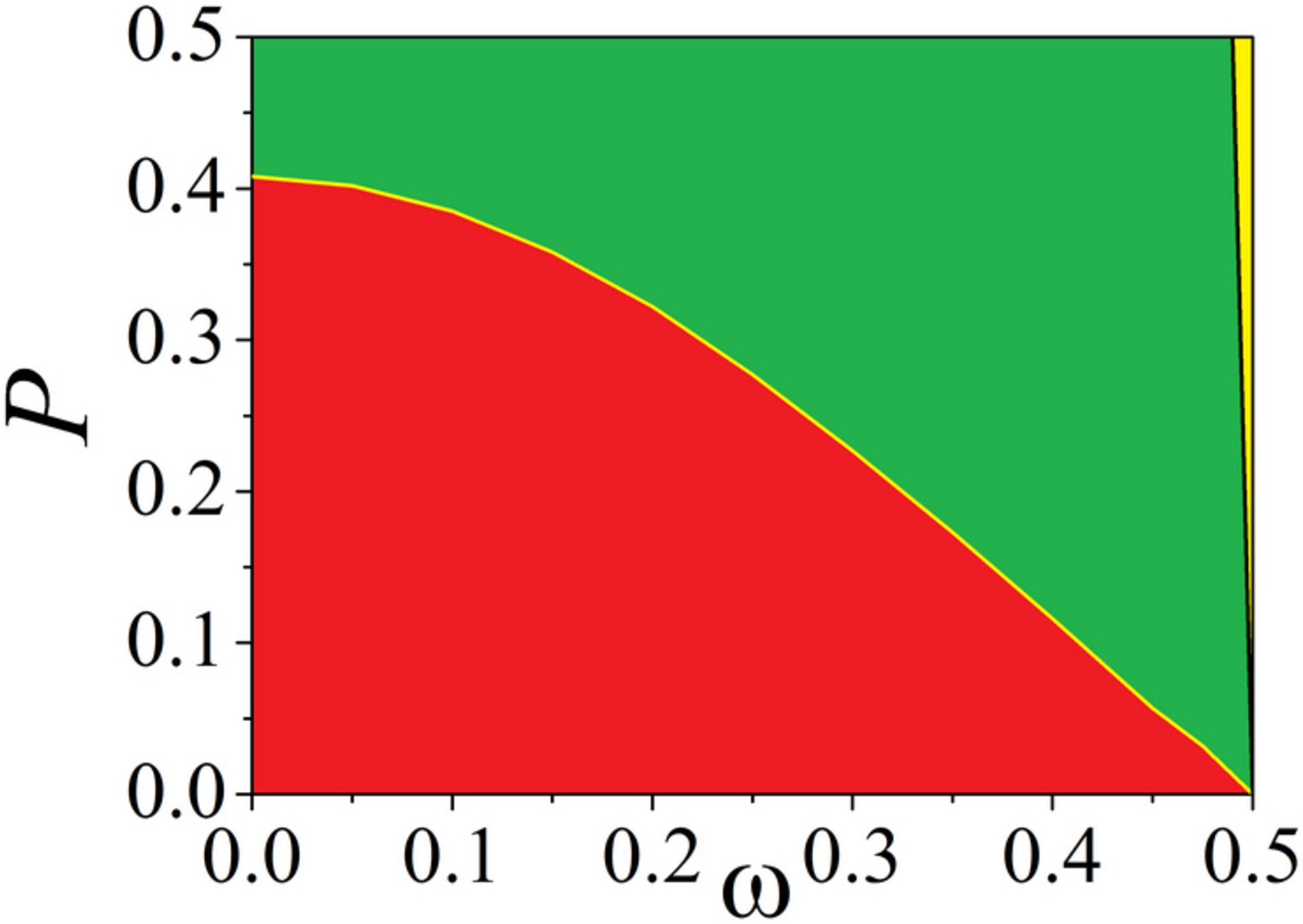}} 
\caption{(Color online) The stability diagrams for the AnS (antisymmetric)
and BAnS (\textit{broken-antisymmetry}) modes in the case of the SDF
nonlinearity for $A=0.5$ (a) and $A=1$ (b). The red (bottom) and green (top)
areas refer to the monostability of the AnS and BAnS modes, respectively.
The bsitability between these species occurs in the yellow (rightmost)
region.}
\label{1stPAomegaSDF}
\end{figure}

Proceeding to the systematic description of the modes supported by the SDF
nonlinearity acting in the combination with the rotating harmonic potential,
we note, first, that the numerical analysis demonstrates the existence and
stability of the mode of the 2HS type at all values of $(\omega ,P,A)$,
therefore this solution is not included in the stability diagrams, which are
displayed for the FHS mode in Fig. \ref{1stPAomegaSDF}, and for the AnS and
newly found BAnS ones in Fig. \ref{AntiPAomegaSDF}.

In the blank area of Fig. \ref{1stPAomegaSDF}, the FHS solutions exist too,
featuring an instability accounted for by a quartet of complex eigenvalues
[see Eq. (\ref{lambda})]. In direct simulations, this instability converts
the stationary mode into an oscillatory one  (not shown here).
%
Further, Fig. \ref{1stPAomegaSDF} shows that the antisymmetry-breaking
transition, which follows the increase of the total power, $P$, is of a
supercritical type. Actually, the supercritical character of the
antisymmetry-breaking bifurcation is a characteristic feature of systems
with the SDF sign of the nonlinearity \cite{Warsaw}.

Finally, the energy comparison for the model with the SDF nonlinearity is
presented in Fig. \ref{fig_9_c}, along the boundary separating the existence
(and stability) regions of the AnS and BAnS modes in Fig. \ref{1stPAomegaSDF}%
(b). The figure shows the energies of these modes, which coincide along the
boundary, along with the results for the FHS and 2HS symmetric modes. From
here, it is concluded that the energies are related as follows:%
\begin{equation}
H_{\mathrm{2HS}}<H_{\mathrm{BAnS}}=H_{\mathrm{AnS}}<H_{\mathrm{FHS}},
\end{equation}%
i.e., the 2HS mode represents the \textit{ground state} in the case of the
SDF nonlinearity, cf. Eqs. (\ref{HHH}) and (\ref{HHHHH}).

\section{CONCLUSION}

The objective of this work is to study the trapped modes of the symmetric,
asymmetric, and antisymmetric types, and the symmetry- or
antisymmetry-breaking transitions between them, in the rotating ring
carrying the DWP (double-well potential) and the cubic nonlinearity with the
SF or SDF (self-focusing or defocusing) sign. The analysis has been
performed in the first rotational Brillouin zone. In the SF case, five types
of different modes, and their stability, have been identified, three
dominated by the fundamental angular harmonic (FHS, FHA, and AnS), and two
others based on the second harmonic, 2HS and 2HA modes. The SSB (spontaneous
symmetry-breaking) transition between the FHS and FHA modes is of the
subcritical type, featuring a conspicuous bistability area, while the
transition between the between the 2HS and 2HA states is supercritical.
There is no overlap between the existence regions of the asymmetric modes of
the FHA and 2HA types, one of which realizes the ground state of the SF
system.

The SDF model supports four distinct species of the trapped modes, \textit{%
viz}., FHS, 2HS, AnS, and, in addition, the BAnS (broken-antisymmetry) mode.
The latter one appears from the AnS state as a result of supercritical
antisymmetry-breaking transition. The 2HS mode represents the ground state
of the SDF system.

The present work can be naturally extended in other directions. First, it is
possible to consider the rotating potential in the form of $A\cos \left(
n\theta ^{\prime }\right) $ with $n\geq 3$, while the present analysis
corresponds to the DWP with $n=2$ [note that Eq. (\ref{V}) suggests a
possibility to create the helical photonic lattice with $n=2S$ and $S>1$].
This generalization will give rise to many new trapped modes; in particular,
it may be possible to build one with a small width, $\Delta \theta ^{\prime
}\ll 2\pi $, hence it may be considered as a \textit{soliton} trapped in the
lattice and rotating along with it, cf. Ref. \cite{HS}. It is relevant too
to consider effects of the two-dimensionality (the entire rotating plane,
rather than a narrow ring). Our preliminary analysis, based on simulations
of the 2D counterpart of Eq. (\ref{Eq5p}), demonstrates essentially the same
types of trapped modes in rotating annuli of a finite radial size, as
reported above in the one-dimensional limit. Another interesting extension
is to replace the linear rotating potential by its nonlinear counterpart,
generated by\ modulation of the local nonlinearity along the angular
coordinate (as was done in Ref. \cite{Pu} in the absence of the rotation).
Results obtained for one-dimensional modes trapped in the rotating nonlinear
potential will be reported elsewhere.

\begin{acknowledgments}
We appreciate help in the use of numerical methods provided by Nir Dror and
Shenhe Fu. This work was supported by Chinese agencies NKBRSF (grant No.
G2010CB923204) and CNNSF(grant No. 11104083,10934011), by the German-Israel
Foundation through grant No. I-1024-2.7/2009, and by the Tel Aviv University
in the framework of the ``matching" scheme.
\end{acknowledgments}

\bibliographystyle{plain}
\bibliography{apssamp}

\end{document}